\documentclass[]{aa}
\usepackage{soul}
\usepackage{natbib}
\def\kms{\hbox{km$\;$s$^{-1}$}}
\def\halpha{H$\mathrm{\alpha}$}
\def\cak{\hbox{\ion{Ca}{ii}~K}}
\def\Mgk{\hbox{\ion{Mg}{ii}~k}}
\def\Si{\hbox{\ion{Si}{IV}}}
\usepackage[dvipsnames]{xcolor}
\usepackage{graphicx}
\usepackage{epstopdf}
\usepackage{lipsum}
\usepackage{mwe}
\usepackage{caption}
\usepackage{subcaption}
\usepackage[varg]{txfonts}
\usepackage[bookmarks = true,colorlinks = true,linkcolor = red,urlcolor = blue,citecolor = blue,pdftex,bookmarks = true,linktocpage = true,hyperindex = true,breaklinks]{hyperref}
\begin{document}

   \title{Evidence of multithermal nature of spicular downflows}

   \subtitle{Impact on solar atmospheric heating}

  \author{Souvik Bose \inst{1}$^,$\inst{2}
  \and
  Luc~Rouppe~van~der~Voort \inst{1}$^,$\inst{2}
  \and
  Jayant~Joshi \inst{1}$^,$\inst{2}
  \and
  Vasco~M.~J.~Henriques\inst{1}$^,$\inst{2}
  \and
  Daniel~N{\'o}brega-Siverio\inst{3}$^,$\inst{4}$^,$\inst{1}$^,$\inst{2}
  \and
  Juan~Mart{\'i}nez-Sykora\inst{5}$^,$\inst{6}$^,$\inst{1}$^,$\inst{2}
  \and
  Bart~De~Pontieu\inst{5}$^,$\inst{1}$^,$\inst{2}
          }

  \institute{Institute of Theoretical Astrophysics, University of Oslo, P.O. Box 1029 Blindern, NO-0315 Oslo, Norway
          \and
    Rosseland Centre for Solar Physics, University of Oslo, P.O. Box 1029 Blindern, NO-0315 Oslo, Norway\\
             \email{souvik.bose@astro.uio.no}
          \and
    Instituto de Astrof\'isica de Canarias, E-38200 La Laguna, Tenerife, Spain
          \and
    Departamento de Astrof\'isica, Universidad de La Laguna, E-38206 La Laguna, Tenerife, Spain
          \and
    Lockheed Martin Solar and Astrophysics Laboratory, Palo Alto, CA 94304, USA
          \and
    Bay Area Environmental Research Institute, NASA Research Park, Moffett Field, CA 94035, USA
           }
 
% \abstract{}{}{}{}{} 
% 5 {} token are mandatory
 \date{\today}
 
  \abstract
  % context heading (optional)
  % {} leave it empty if necessary  
   {Spectroscopic observations of the emission lines formed in the solar transition region commonly show persistent downflows of the order of 10--15~\kms{}. The cause of such downflows, however, is still not fully clear and has remained a matter of debate. }
  % aims heading (mandatory)
   {We aim to understand the cause of such downflows by studying the coronal and transition region responses to the recently reported chromospheric downflowing rapid red shifted excursions (RREs), and their impact on heating the solar atmosphere.}
  % methods heading (mandatory)
   {We have used two sets of coordinated data from the Swedish 1-m Solar Telescope, the Interface Region Imaging Spectrograph, and the Solar Dynamics Observatory for analyzing the response of the downflowing RREs in the transition region and corona. To provide theoretical support, we use an already existing 2.5D magnetohydrodynamic simulation of spicules performed with the Bifrost code. }
  % results heading (mandatory)
   {We find ample occurrences of downflowing RREs and show several examples of their spatio-temporal evolution, sampling multiple wavelength channels ranging from the cooler chromospheric to hotter coronal channels. These downflowing features are thought to be likely associated with the returning components of the previously heated spicular plasma. Furthermore, the transition region Doppler shifts associated with them are close to the average red shifts observed in this region which further implies that these flows could (partly) be responsible for the persistent downflows observed in the transition region. We also propose two mechanisms (a typical upflow followed by a downflow and downflows along a loop), from the perspective of numerical simulation, that could explain the ubiquitous occurrence of such downflows. A detailed comparison between the synthetic and observed spectral characteristics, reveals a distinctive match, and further suggests an impact on the heating of the solar atmosphere.}
  % conclusions heading (optional), leave it empty if necessary 
   {We present evidence that suggests that at least some of the downflowing RREs are the chromospheric counterparts of the transition region and lower coronal downflows. }
% abstract: limit to 300 words, and be self-contained %%297 words

 \authorrunning{Bose et al.}

\maketitle
%

%-------------------------------------------------------------------

\section{Introduction}
\label{Section:Introduction}

The emission spectral lines formed in the upper atmosphere of the Sun, in particular the transition region, are on average found to be persistently Doppler shifted towards the redward side of the respective line centers. In other words, these layers, existing at temperatures ranging between 0.05~MK to \textasciitilde0.25~MK, show abundant downward flows that appear to last from several hours to (sometimes) several days \citep[e.g.][]{1976ApJ...205L.177D,1981ApJ...251L.115G,1982SoPh...77...77D}. Decades of observations, since the launch of NASA's \textit{Skylab} missions in the 1970s, have revealed this puzzling phenomenon in the Sun \citep{1976ApJ...205L.177D,1981ApJ...251L.115G,1982SoPh...77...77D,1988ApJ...334.1066K,1997SoPh..175..349B,1999ApJ...516..490P,1999ApJ...522.1148P,2011A&A...534A..90D}. These studies have shown that the average line profiles formed in the transition region are red shifted by as much as 10--15~\kms{}, implying the presence of wave motions or plasma flows with amplitudes that are considerable fractions of the speed of sound. Persistent red shifts have also been observed in the spectra of late type stars, indicating that this phenomenon is rather common not just in the Sun but also (at least) among the relatively cool stars \citep[see][]{1996ApJ...458..761W,2004A&A...415..331P}.

Later, observations from SoHO and Hinode's Extreme ultraviolet Imaging Spectrograph, have revealed that one can often find a net upflow or blue shift of the order of a few \kms{} associated with the spectral lines that are sensitive beyond \textasciitilde1~MK (coronal lines), when observed close to the disk-center \citep{1999ApJ...516..490P,1999ApJ...522.1148P,2009ApJ...701L...1D,2012ApJ...749...60M}. These studies indicate that there could be a temperature dependence of the Doppler shift of these spectral lines, with a transition from red to blue shifts. \citet{2012ApJ...749...60M} suggested that this conversion takes place roughly around 1~MK. This transition provides an important constraint in distinguishing the different models of the upper solar atmosphere, in particular the transition region.

The above studies have led to the development of several hypotheses, both from the perspective of observations and numerical simulations, to explain the observed shifts of the coronal and transition region lines. Several theories, such as downward propagating acoustic waves generated by nanoflares in the corona \citep{1993ApJ...402..741H}, return of the previously heated spicular material \citep{1977A&A....55..305P,1982ApJ...255..743A}, rapid episodic heating at the base of the corona \citep{2010ApJ...718.1070H}, and downward propagating pressure disturbances \citep{2018A&A...614A.110Z}, have been proposed in the past but no definitive consensus has emerged so far. 

Spicules, being ubiquitous, could play an important role in explaining these observed Doppler shifts in the upper atmosphere. Since the average upward mass flux carried by spicules is roughly 100 times more than what is needed to contribute towards the solar wind, \citet{1977A&A....55..305P} proposed for the first time that the remaining heated spicular plasma undergo cooling, and is eventually drained back to the chromosphere via the transition region. These downflowing spicular materials were therefore conjectured to be responsible for the observed red shifts in the transition region lines. Later \citet{1982ApJ...255..743A} and \citet{1984ApJ...287..412A} also supported this scenario based on their extensive modeling efforts. However, there has been a lack of direct observational signatures, and more advanced theoretical models have not been able to reproduce these initial observations \citep{1987ApJ...319..465M}. Nonetheless, recent extreme ultraviolet (EUV) and UV observations from high-resolution space-based telescopes have revived massive interest in the role of spicules in mass and energy supply to the upper atmospheres of the Sun \citep{2009ApJ...701L...1D,2009ApJ...707..524M,2011Sci...331...55D}, with \citet{2012ApJ...749...60M} finding strong observational indications of both lower coronal and transition region flows attributed to spicules. Interestingly, \citet{2012ApJ...749...60M} also found signatures of downflowing patterns in the relatively cooler emission channels of the lower solar corona that were speculated to be returning heated spicule plasma. In fact, the observations of \citet{2009ApJ...707..524M} suggest that the high-speed upflows are associated with discrete mass-injection events caused by rapid episodic heating in the solar chromosphere in the form of spicules. This proposition shows some similarities to the explanation provided by \citet{2010ApJ...718.1070H}, who observed similar episodic heating in their numerical simulations which further led to the production of bi-directional flows in the interface between the transition region and lower solar corona.

Despite abundant observations of such downflows, it is not very clear as to why their signatures were not prominent in the deeper layers of the Sun, such as the chromosphere. An obvious hypothesis could be attributed to the enormous difference in the plasma densities between the deeper and the upper layers (i.e. the chromosphere and the transition region), which might cause these downflows to disintegrate as soon as they reach chromospheric heights. However, recent observations and analysis with high-resolution datasets acquired from the Swedish 1-m Solar Telescope \citep[SST,][]{2003SPIE.4853..341S} by \citet{2021A&A...647A.147B}, show that rapid downflows in the form of chromospheric spicules are observed in abundance when viewed in the red wing images of the \halpha{} spectral line. These downflows, termed as downflowing rapid red-shifted excursions (downflowing RREs), were found to have an apparent motion that is opposite to their traditional counterparts, namely the rapid blue and red shifted excursions (RBEs/RREs). \citet{2008ApJ...679L.167L}, and later \citet{Luc_2009} observed the spectral signatures of RBEs in high-resolution \ion{Ca}{II}~8542~\AA\ and \halpha{} datasets, and linked them to the on-disk counterparts of type-II spicules that were first observed (off the limb) by \citet{Bart_2007_PASJ}. The RREs, first reported by \citet{Bart_3_motions}, were found to be the red wing counterparts of RBEs, and their occurrence was attributed  to the complex twisting and swaying motions exhibited by type-II spicules. Both RBEs and RREs are associated with upflows along the chromospheric magnetic field lines, and are thought to be a manifestation of the same phenomenon. The occurrence of the latter was attributed to the complex torsional and transverse motions associated with RBEs, that sometimes resulted in a net red shift when observed on the disk \citep[see discussions by][]{Bart_3_motions,2013ApJ...769...44S,Bart2014,2015ApJ...802...26K}.

The downflowing RREs, however, which are retracting rather than expanding and moving away from the magnetic network, were associated with downward mass motions along the magnetic field lines, and were speculated to be the (much sought after) chromospheric representatives of the transition region and lower coronal downflows \citep{2021A&A...647A.147B}. Even though, type-II spicules have been shown to exhibit typical parabolic up-down behavior with lifetimes around 600~s \citep{Tiago_2014_heat,2019Sci...366..890S}, their impact on the net red shifts of the transition region lines and their connection to the chromospheric spicular counterparts have not yet been studied in great detail. In this paper, using two sets of coordinated ground and space-based datasets with SST, the Interface Region Imaging Spectrograph \citep[IRIS,][]{Bart2014}, and the Atmospheric Imaging Assembly \citep[AIA,][]{2012SoPh..275...17L} on-board the Solar Dynamics Observatory \citep[SDO,][]{2012SoPh..275....3P}, we show that many chromospheric spicular downflows show signatures in the transition region and the lower corona. Focusing on enhanced network and quiet Sun targets close to the disk center, we find that these downflows are ubiquitous and multithermal in nature with many of them being heated to temperatures as high as 0.8~MK. Analysis, based on their spatio-temporal evolution and spectral signatures, indicate that these downflows are associated with real mass motions. Furthermore, a comparison with a state-of-the-art magnetohydrodynamic (MHD) spicule simulation reveals a good match with the observations and further indicates that these flows show an enhancement in temperature and undergo increased heating during the downflowing phase. This further affects the formation of emission lines in the upper chromosphere, that in turn could impact the heating of the solar chromosphere.

The outline of the paper is structured as follows. Section~\ref{Section:obs_simulations} describes the observations used in this study and provides an overview of the numerical simulation. Section~\ref{section:method} describes the methodology used for the identification of the rapid downflows in the observations, followed by a description of the radiative transfer techniques used. This is followed by Sect.~\ref{Section:results}, that described the results of the analysis performed, and finally, the discussions and conclusions are presented in Sects.~\ref{Section:Discussions} and \ref{Section:Conclusion}, respectively. 

%--------------------------------------------------------------------
\section{Observations and numerical simulation}
\label{Section:obs_simulations}
\subsection{Observations and data reduction}
\label{Subsection:Obs}

\begin{figure*}[!ht]
\centering
\includegraphics[bb = 20 0 1278 320, width=\textwidth]{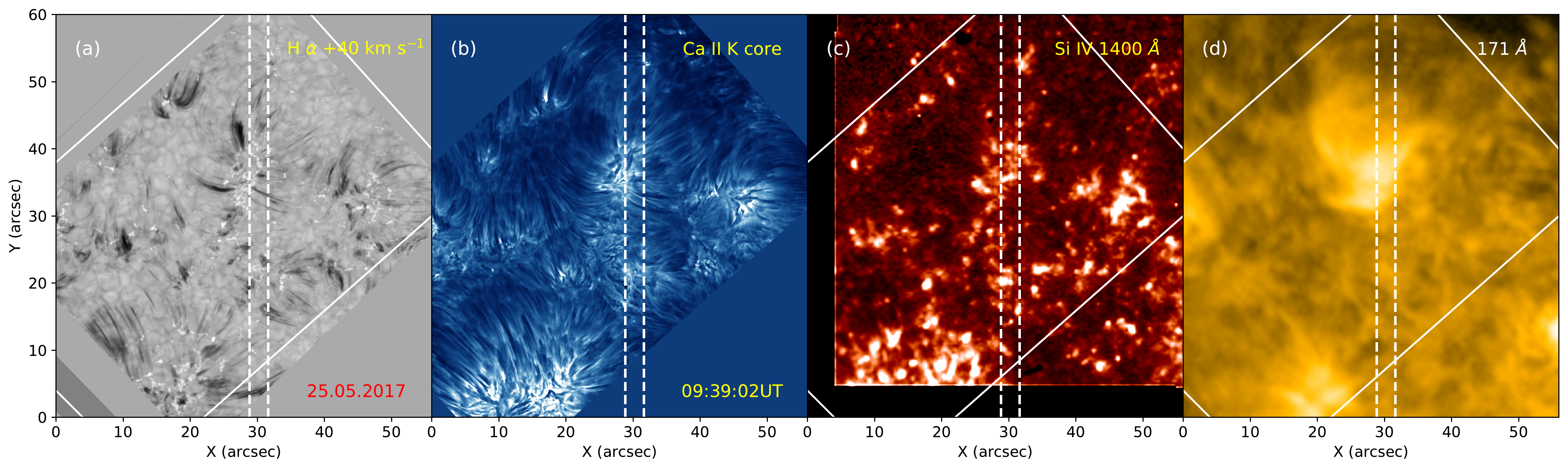}
\includegraphics[bb = 20 0 1680 450, width=\textwidth]{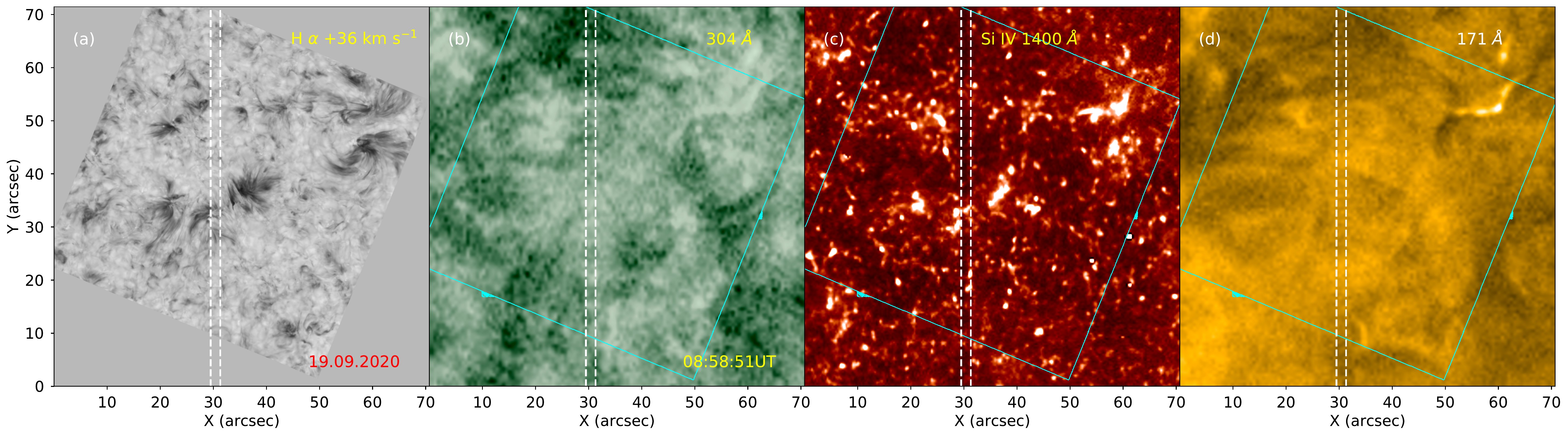}
\caption{\label{figure:Context_obs}%
Full FOV of the observations corresponding to datasets~1 (top row) and 2 (bottom row). Top row: panel~a shows the \halpha{} red wing image observed with CRISP at a Doppler offset of +40~\kms{}, panel~b shows the corresponding CHROMIS \cak{} line core image, panels~c and d correspond to co-spatial \Si{}~1400~\AA\ SJI from IRIS and 171~\AA\ image from AIA, respectively. Bottom row: panel~a shows the \halpha{} red wing image observed at a Doppler offset of +36~\kms{} acquired with CRISP, and panels~b--d show the corresponding co-spatial cutouts from AIA 304~\AA, IRIS \Si{}~1400~\AA\ SJI, and AIA 171~\AA, respectively. The dashed vertical lines correspond to the spatial extent of the IRIS rasters in the two datasets and the direction to solar north is pointing upward. The white and cyan boxes in the top and bottom rows represent the area of the \ion{Ca}{ii}~K core and \halpha{} red wing images, respectively. Animations corresponding to these figures are available at \url{https://www.dropbox.com/sh/k5vwkp49syf35v3/AAA8V7WC_5Mxjl0GKMOCGYhEa?dl=0}.
}
\end{figure*}

\begin{figure*}
   \centering
   \includegraphics[width=\textwidth]{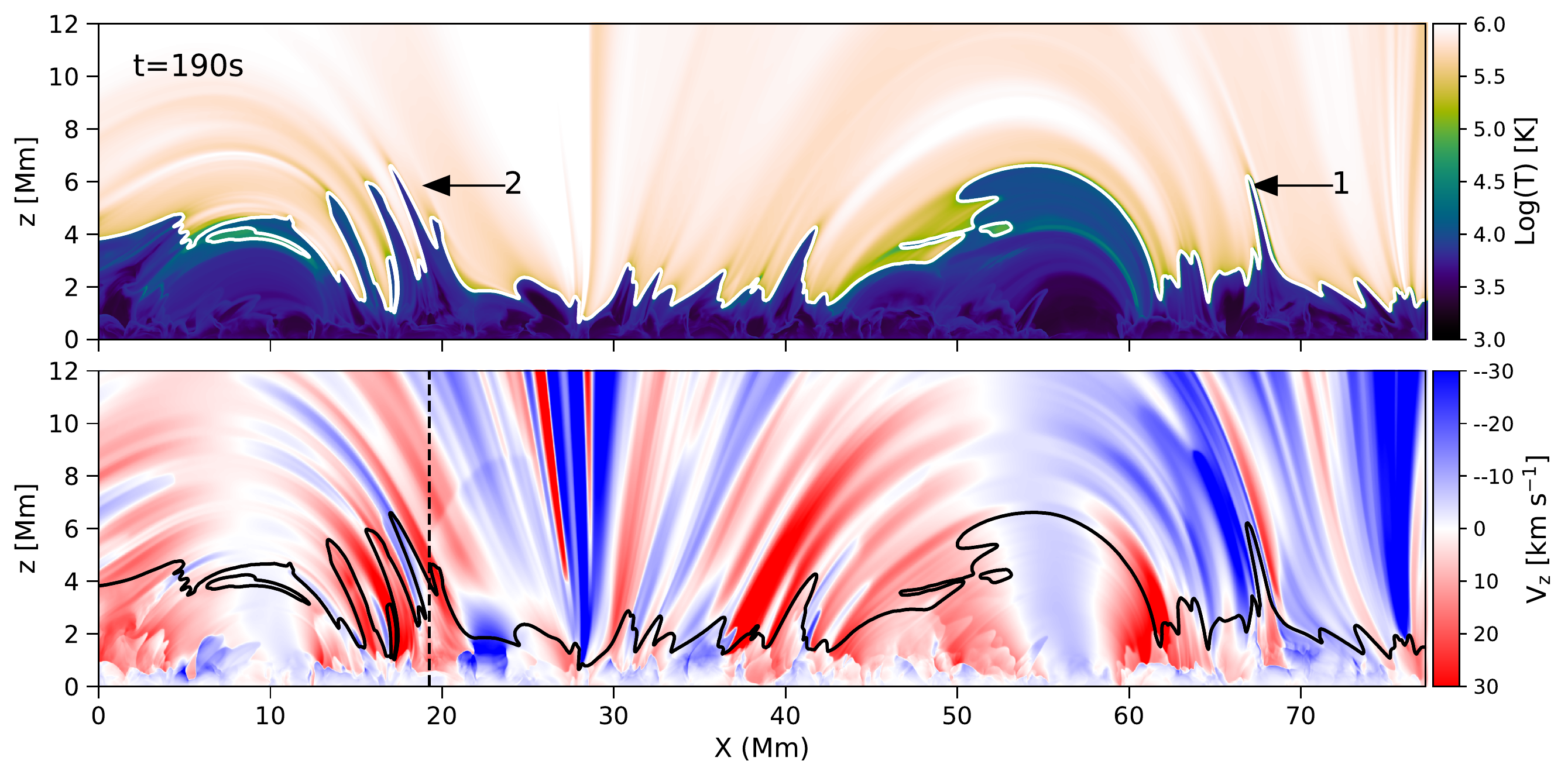}
   \caption{Overview of the MHD simulation snapshot from \citet{Juan_2017_Science} showing a 2D slice of temperature (top row) and signed vertical velocity v$_{\mathrm{z}}$ (bottom row) at $t$=190~s from the start. The spicules appear as cool finger-like intrusions in the hot corona and the ones which are further analyzed are numbered as 1 and 2. Velocities away from the observer are indicated in red (positive) while towards the observer are indicated in blue (negative). The contour indicates the location where the temperature equals 30~kK. The region to left of the dashed vertical line in the bottom row indicates the region-of-interest that has been further studied in Sect.~\ref{subsection:discussion_loop_drre}. An animation of this figure is available at \url{https://www.dropbox.com/s/ee67bkpsye7v01h/Context_sim_movie.mp4?dl=0}.
   } 
        \label{figure:Context_v1_sim}%
    \end{figure*}

For this study, we analyzed two datasets from coordinated SST and IRIS campaigns, on 25 May 2017 (henceforth dataset~1) and on 19 September 2020 (henceforth dataset~2). Dataset~1 focused on an enhanced network region close to the disk-center with solar ($X$,$Y$) = (45\arcsec, $-93$\arcsec), and a duration close to 97 min. The CRisp Imaging Spectropolarimeter \citep[CRISP,][]{Crisp_2008} was used to acquire imaging spectroscopic and spectropolarimetric data in \halpha{} and \ion{Fe}{I}~6302~\AA\ respectively, while the CHROmospheric Imaging Spectrometer (CHROMIS) was used to record imaging spectroscopy data in \cak{}. Moreover, IRIS was running in a medium-dense 8 step raster mode (OBS-ID \verb|3633105426|) with a 2~s exposure time and spectrograph slit covering a field-of-view (FOV) of 2\farcs8 and 62\farcs0 in the solar $X$ and $Y$ directions. The IRIS dataset was expanded up to CHROMIS pixel scale (0\farcs038), and was co-aligned by cross-correlating the inner wings of \cak{} with 2796~\AA\ slit-jaw images (SJI) such that the cadence of the co-aligned sequence is roughly 13.6~s. The common FOV of the co-aligned sequences was roughly 56\arcsec\ $\times$ 60\arcsec\ in the solar $X$ and $Y$ directions. The top row of Fig.~\ref{figure:Context_obs} shows a scan from this co-aligned dataset with panels a and b showing SST CRISP (\halpha{}) red wing image at +40~\kms{} and CHROMIS (\cak{}) line core image, whereas panels c and d show the corresponding IRIS \Si{} 1400~\AA\ SJI and AIA 171~\AA\ images. 
%This dataset has been used (without the co-aligned SDO sequences) in two former publications by the first author, and we refer the reader to \citet{my_paper_3} and \citet{2021A&A...647A.147B} for further details on these observations such as, data reduction and precise co-alignment techniques.
This dataset has been used (without the co-aligned SDO sequences) by \citet{my_paper_3} and \citet{2021A&A...647A.147B} where further details on these observations such as data reduction and precise co-alignment techniques can be found.

Dataset~2 focused on a quiet Sun region at the disk center located around solar ($X$,$Y$) = ($-1$\arcsec, 1\arcsec) shown in the bottom row of Fig.~\ref{figure:Context_obs}. The duration of this data was 59~min starting from 08:23~UT and lasting until 09:22~UT. CRISP recorded imaging spectroscopic data in \halpha{} and \ion{Ca}{II}~8542~\AA\, in addition to spectropolarimetric data in \ion{Fe}{I}~6173~\AA. The \halpha{} spectral line was sampled at 31 wavelength positions between $\pm$~1.5~\AA, in steps of 100~m\AA, \ion{Ca}{II}~8542~\AA\ was sampled at 9 wavelength positions between $\pm$~0.8~\AA\, in steps of 200~m\AA\,, and \ion{Fe}{I}~6173~\AA\ was sampled at 14 wavelength positions between $-$0.32~\AA\ and +0.68~\AA\ with respect to the line cores. The temporal cadence of the time series was 23.6~s with a spatial sampling of 0\farcs058. High spatial resolution down to the diffraction limit of the telescope (given by $\lambda/\mathrm{D}$ = 0\farcs16 at 6563~\AA) was achieved through high-quality seeing conditions, the SST adaptive optics system \citep{2019A&A...626A..55S} and the Multi-Object Multi-Frame Blind Deconvolution (MOMFBD) technique described in \citet{vannoort2005MOMFBD}. 
%We used the CRISPRED reduction pipeline described in \citet{2015ApJ...810..145D} 
We used the SSTRED reduction pipeline \citep{2015ApJ...810..145D,
2018arXiv180403030L} to facilitate further reduction of the data, while including the spectral consistency technique outlined in \citet{2012A&A...548A.114H}. Furthermore, intra-CRISP alignment between \halpha{} and \ion{Ca}{II}~8542~\AA\ was achieved by cross-correlating the corresponding photospheric wideband images, followed by a de-stretching procedure to compensate the warping due to seeing effects. In this study we only focus on the spicular downflows observed in the red wings of \halpha{}.  

IRIS ran in a large-dense 4-step raster mode (OBS-ID \verb|3633109417|), co-observing the same FOV as SST. The raster cadence was 36~s (with a step cadence of 9~s and an exposure time of 8~s per slit position) and the FOV covered by the spectrograph slit was 1\farcs36 $\times$ 120\arcsec\ in the solar $X$ and $Y$ direction. Moreover, it also recorded slit-jaw images (SJIs) in the 2796~\AA\ channel (that is dominated by the inner wings and core of the \Mgk{} spectral line), and \Si{} 1400~\AA\ channel with a cadence of 18~s covering a FOV of 120\arcsec $\times$ 120\arcsec. The high exposure time (in comparison to dataset~1) of this dataset enabled better visibility of the different features in the \Si{} channel.

The IRIS data were co-aligned to SST by cross-correlating \ion{Ca}{II}~8542~\AA\ data (integrated over the entire wavelength dimension) with the SJI recorded in the 2796~\AA\ channel (expanded to the CRISP pixel scale of 0\farcs058). The summation over the wavelength resulted in images that bore a close resemblance to the features observed in the 2796~\AA\ channel thereby enabling better correlation. The resulting IRIS data was finally cropped so as to have a common FOV of 70\arcsec $\times$ 70\arcsec with SST. Panels a, and c in the bottom row of Fig.~\ref{figure:Context_obs} show the co-aligned SST CRISP image observed in the red wing (+36~\kms{}) of \halpha{} and the corresponding \Si{}~1400~\AA\ SJI, respectively. The coordinated datasets (both 1 and 2) will be made publicly available in the future as a part of the SST--IRIS database \citep{2020A&A...641A.146R}.

\paragraph{Observations from Solar Dynamics Observatory (SDO)}
For both dataset~1 and dataset~2, we downloaded corresponding SDO image cut-out sequences. These sequences were cross-aligned (all AIA channels to Helioseismic Magnetic Imager (HMI) continuum), and then the HMI continuum images were finally co-aligned to \cak{} and \ion{Ca}{II}~8542~\AA\ wideband channels for dataset~1 and 2, respectively. We used an Interactive Data Language (IDL) based automated aligning procedure developed by Rob Rutten for this purpose, which is publicly available at his website\footnote{\url{https://webspace.science.uu.nl/~rutte101/rridl/00-README/sdo-manual.html}}. For a precise cross-alignment among the SDO channels, the procedure collects and downloads two sets of SDO cutouts -- one at full cadence focusing on the smaller target area (SST wideband) and the other large ones (700~\arcsec~$\times$~700~\arcsec) around disk-center at a lower cadence. The latter set is used to find offsets between the different channels by cross-correlation of smaller sub-fields covering roughly 30\arcsec~$\times$~30~\arcsec, and by applying the height-of-formation differences in an iterative fashion. Since both the datasets are close to the disk center, the projection effects are likely lower, which suggests that the accuracy of cross-alignment could be on the order of an AIA pixel (0\farcs6) or possibly less, depending on the overall magnetic topology. We refer the reader to \citet{2020arXiv200900376R} for more details. Each co-aligned SDO dataset comprises eleven (nine AIA and two HMI) image cutout sequences that are resampled from their original pixel scale (0\farcs5) to CHROMIS and CRISP pixel scale of 0\farcs038 and 0\farcs058 respectively for the two datasets, and are chosen such that they are as close as possible in time (through nearest-neighbour selection) with respect to one another. Sample co-aligned SDO AIA images are shown in panel d (top row) and panels b and d (bottom row) of Fig.~\ref{figure:Context_obs}.  For the purpose of the analysis presented in this paper, we use only the EUV AIA 304~\AA, 171~\AA, and 131~\AA\ sequences.

\subsection{Numerical simulation}
\label{subsection:numerical_sim}

% For the purpose of comparing the observed properties of the downflowing RREs, we use the state-of-the-art 2.5--dimensional (2.5D) radiative magnetohydrodynamic (MHD) numerical simulation of spicules simulated using the \verb|Bifrost| code \citep{2011A&A...531A.154G}, and was described for the first time by \citet{Juan_2017_Science}. Contrary to older simulations, thin, needle, spicule-like features were produced naturally in this case and they occurred ubiquitously throughout the simulation domain. The importance of ambipolar diffusion \citep{2020A&A...638A..79N} along with other complex physical mechanisms such as, ion-neutral interaction effects in the partially ionized plasma, large-scale magnetic fields (that is, magnetic loops with length \textasciitilde 50~Mm) and high spatial resolution (< 30~km) were found to be crucial for the observed spicular ubiquity. Furthermore, this model also successfully reproduced various observed properties of type-II spicules such as strongly collimated flows (\textasciitilde100~\kms{}) that reach up to 10~Mm in height within a lifetime of 2--10~min. 
% \if 1 > 2
For the purpose of comparing the observed properties of the downflowing RREs, we use 2.5--dimensional (2.5D) radiative magnetohydrodynamic (MHD) numerical simulation of spicules from \citet{2017ApJ...847...36M,Juan_2017_Science}. This simulation was performed using the state-of-the-art Bifrost code \citep{2011A&A...531A.154G}, including ambipolar diffusion\footnote{The latest module for the implementation of ambipolar diffusion in the Bifrost code can be found in \citet{2020A&A...638A..79N}.} to model spicule generation.
Contrary to older Bifrost simulations, high-speed (\textasciitilde50--100~\kms{}), thin, needle, spicule-like features were produced naturally in this case and they occurred ubiquitously throughout the simulation domain. The importance of ambipolar diffusion along with large-scale magnetic fields (that is, magnetic loops with length \textasciitilde 50~Mm) and high spatial resolution were found to be crucial for the simulated spicular ubiquity. Furthermore, this model also successfully reproduced various observed properties of type-II spicules such as strongly collimated flows (\textasciitilde100~\kms{}) that reach up to 10~Mm in height within a lifetime of up to ~\textasciitilde10~min. 
% \fi 
%
%It also predicted that most of the spicular activity is concentrated 
Most of the spicular activity is concentrated around network and plage regions which is also compatible with observations. %Naturally, this formed a fair choice for the purpose of comparison with the observations.

The simulation domain spans from 2.5~Mm below the visible photosphere up to about 40~Mm into the corona with resolution varying from 12~km to 60~km in the vertical direction. The horizontal domain spans 96~Mm with a uniform resolution of \textasciitilde12~km. Figure~\ref{figure:Context_v1_sim} shows an overview of the simulation snapshot with the top row showing a 2D map of temperature and the bottom row showing the corresponding vertical velocity. The type-II spicules appear as cooler chromospheric intrusions into the hot corona. For the sake of visualization, the vertical domain has been limited to only 12~Mm. The complex physical processes that lead to the generation of spicules, along with other details of the simulation setup has been described extensively both in \citet{2017ApJ...847...36M,Juan_2017_Science} and \citet{Juan_2018}.

\section{Method of analysis}
\label{section:method}

\subsection{Identification of the downflowing spicular events}
\label{subsection:detection}

The downflowing RREs, like RREs, show significant absorption asymmetries in the red wing of \halpha{} spectral line. This formed the basis of an automated detection algorithm based on the $k$-means clustering technique, that enabled the detection of tens of thousands of spicular events including downflowing RREs, RBEs and RREs in dataset~1. The details of the algorithm along with the advanced morphological processing techniques have been described in \citet{my_paper_3}, and more recently in \citet{2021A&A...647A.147B}. In this paper, we simply use those detections and overlay them on the co-aligned IRIS and SDO datasets to study their responses in the transition region and the solar corona. 

For dataset~2, we relied on a detection method that is based on the morphology of the events observed in the far red wing (+36~\kms{}) of the \halpha{} spectral line, as shown in the bottom row of Fig.~\ref{figure:Context_obs}. The events were selected by constructing \halpha{} Doppler maps (blue -- red wing image subtraction at $\pm$~36~\kms{}) where the RBEs show up as dark streaks, while the RREs/downflowing RREs show up as bright ones. Simple thresholding based on intensity of the resulting Doppler maps, led to the extraction of spicular features in the red wing of \halpha{}. This approach has been used in the past to detect RBEs and RREs, and we refer to \citet{Luc_2009,Luc_2015} and \citet{Vasco_2016} for more details. We relied on a simpler detection method for dataset~2 because the major goal was to simply extract the location of the downflowing RREs so that we could investigate their responses in hotter passbands on an event-by-event basis, rather than focusing on their detailed statistical characterization. Moreover, the idea was also to highlight that the downflowing RREs are equally ubiquitous in regions with different magnetic field environments. These detections were further confirmed by extensive visualization using CRISPEX \citep{2012ApJ...750...22V}, a widget-based analysis tool written in IDL. The animations associated with Fig.~\ref{figure:Context_obs} clearly convey that both the datasets are replete with downflowing RREs (in the chromosphere), whose apparent motion is directed towards the strong network regions, opposite to what we typically observe in RBEs and RREs \citep[see][and references therein]{Luc_2009,Bart_3_motions,Sekse_2012,2013ApJ...769...44S,2015ApJ...802...26K}. 

In both datasets~1 and 2, the detected events were then simply overlaid on the corresponding co-aligned IRIS and SDO datasets to study their responses in the transition region and the solar corona. To aid visibility, we subtracted an average over the whole time series from each AIA image on a per-channel basis. This is done separately for 131~\AA, 171~\AA, and 304~\AA\ channels which ensured the removal of the effect of long-lived brighter structures (e.g., loops) along the enormous line-of-sight (LOS) superposition in an optically thin plasma. Finally, the events, now observed across the different wavelengths (from SST, IRIS, and SDO), were again analyzed and visualized using CRISPEX. The evolution of the downflowing RREs was highlighted by constructing space-time ($X$-$t$) maps for each event that are generated by computing the average over the highlighted regions for each time step.

\subsection{Optically thick and thin radiative transfer}
\label{subsection:RH1.5D}

We have used the RH1.5D\footnote{The code is publicly available at :\url{https://github.com/ITA-Solar/rh}} \citep{Tiago_RH_2015} radiative transfer code for optically thick radiative transfer calculations. RH1.5D is a massively parallelized version of the former RH radiative transfer code by \citet{Uitenbroek_2001}. It is capable of performing multi-atom, multi-level non-local thermodynamic equilibrium (LTE) radiative transfer computations under both partial and complete frequency redistribution (PRD and CRD). Moreover, it can also synthesize full Stokes parameters while including Zeeman splitting effects. This code can compute synthetic spectra from either 3D, 2D, or 1D model atmospheres on a column-by-column (1.5D) basis. It also scales well to at least thousands of CPU cores and can be easily run over multiple nodes in a supercomputing cluster. 

We used the RH1.5D code to generate synthetic \cak{} and \Mgk{} spectra from the 2.5D numerical simulation described in Sect.~\ref{subsection:numerical_sim}. The goal of this analysis is to compare the spectral properties of type-II spicules from the simulation with available observations from IRIS and SST. Such an investigation leads to a detailed understanding of spectral line formation in stellar atmospheres. We used a five-bound-level plus continuum model of the \ion{Ca}{II} ion, as used by \citet{2018A&A...611A..62B} and a ten-level plus continuum model of the \ion{Mg}{II} atom used by, e.g. \citet{2013ApJ...772...90L} and \citet{Bose_2019}, under PRD and 1.5D geometry. The fast angle hybrid approximation scheme was used for PRD calculations as described in \citet{2012A&A...543A.109L}. The initial level populations were estimated by assuming a zero mean radiation field and solving the statistical equilibrium equations. This approach works better than starting with LTE populations, for instance, since atoms with strong lines have level populations that are significantly different from LTE \citep{Tiago_RH_2015}. The effects of PRD are important in the region where the K$_{2}$/k$_{2}$ emission peaks are formed since radiative damping is much larger than collisional damping \citep{2017A&A...597A..46S,2018A&A...611A..62B}. Furthermore, these emission peaks along with the wings of resonance lines under consideration, such as \cak{} and \Mgk{}, are formed deeper in the solar atmosphere in comparison to the line core, where the effect of horizontal radiative transfer is small, thereby justifying the 1.5D treatment. The computed specific intensity is further spectrally and spatially smeared with the SST and IRIS instrumental point spread functions to compare with the observations. 

The analysis of synthetic intensity corresponding to the \Si{}~1393~\AA\ spectral line is computed assuming optically thin approximation under ionization equilibrium of \Si{}, in a way similar to \citet{2016ApJ...817...46M}, for example. We use the \verb|CHIANTI| database \citep{2009A&A...498..915D} to synthesize the plasma emission with the ionization balance available in the \verb|CHIANTI| distribution and using the  \Si{} abundance from \citet{2009ARA&A..47..481A} -- more specifically, 7.52 in the customary astronomical scale, where 12 corresponds to hydrogen. The resulting emission spectra was convolved with the IRIS spectral transmission function before comparing with the observations.

\section{Results}
\label{Section:results}
\begin{figure*}[!h]
\centering
\includegraphics[bb = 0 15 690 310, width=0.90\textwidth]{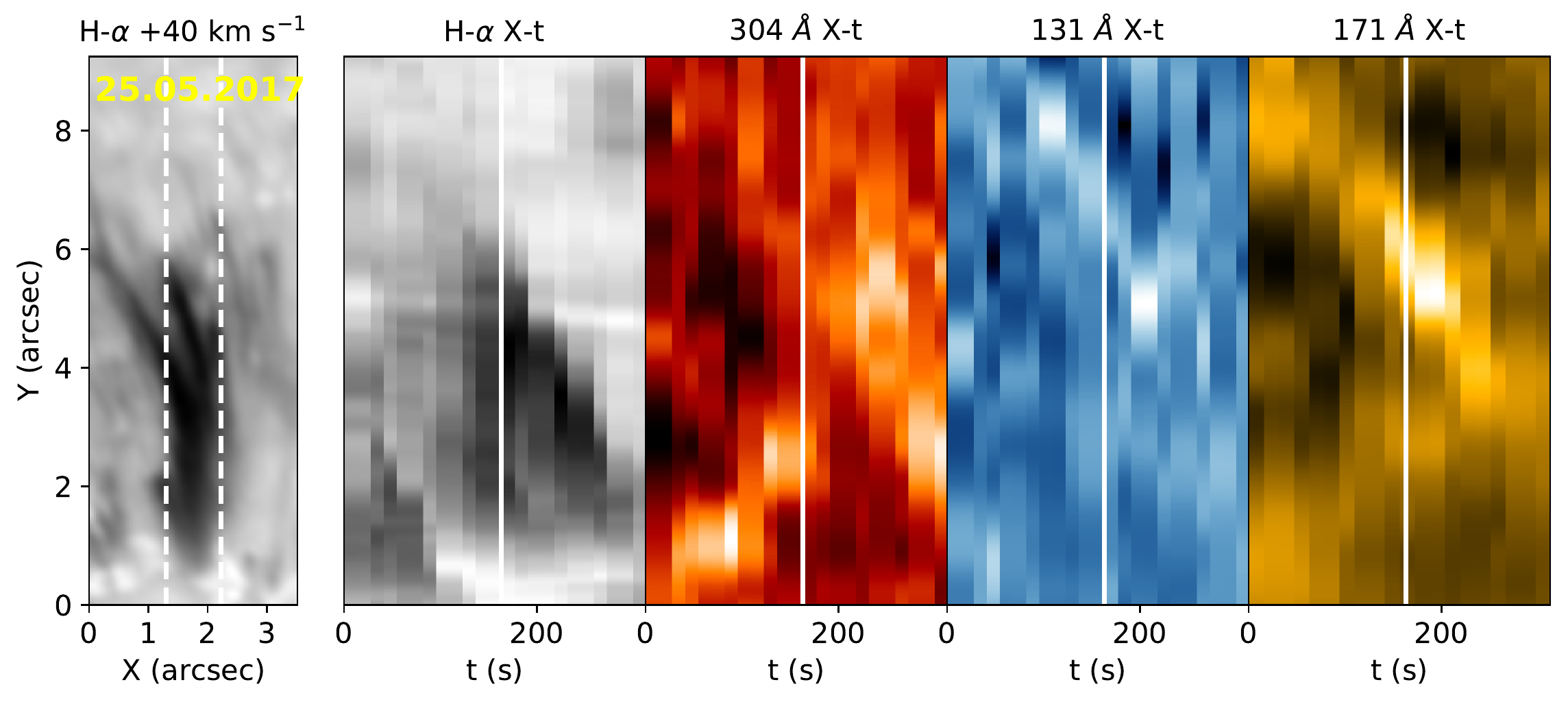}
\includegraphics[bb = 0 15 698 180, width=0.90\textwidth]{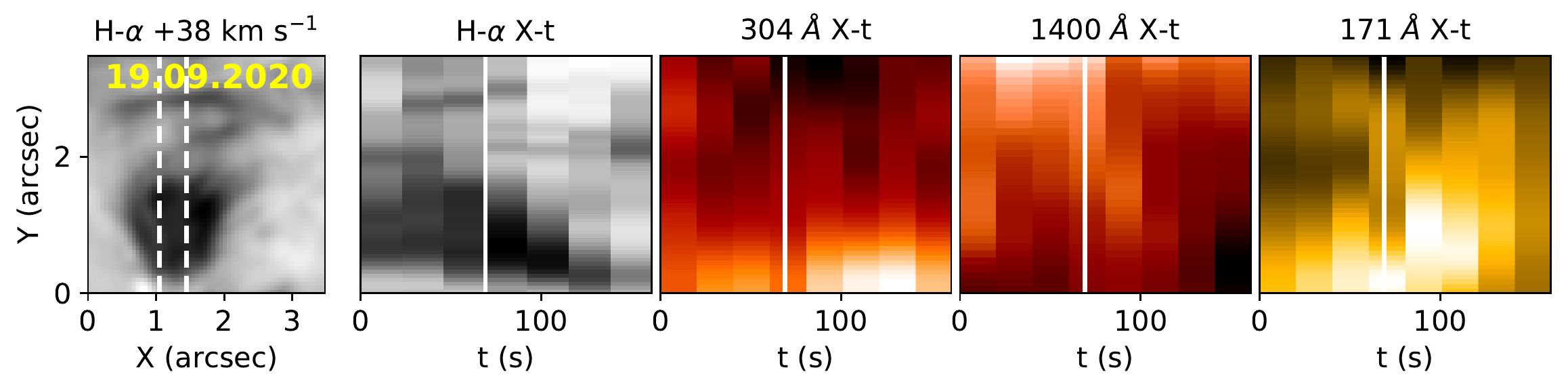}
\includegraphics[bb = 0 15 696 310, width=0.90\textwidth]{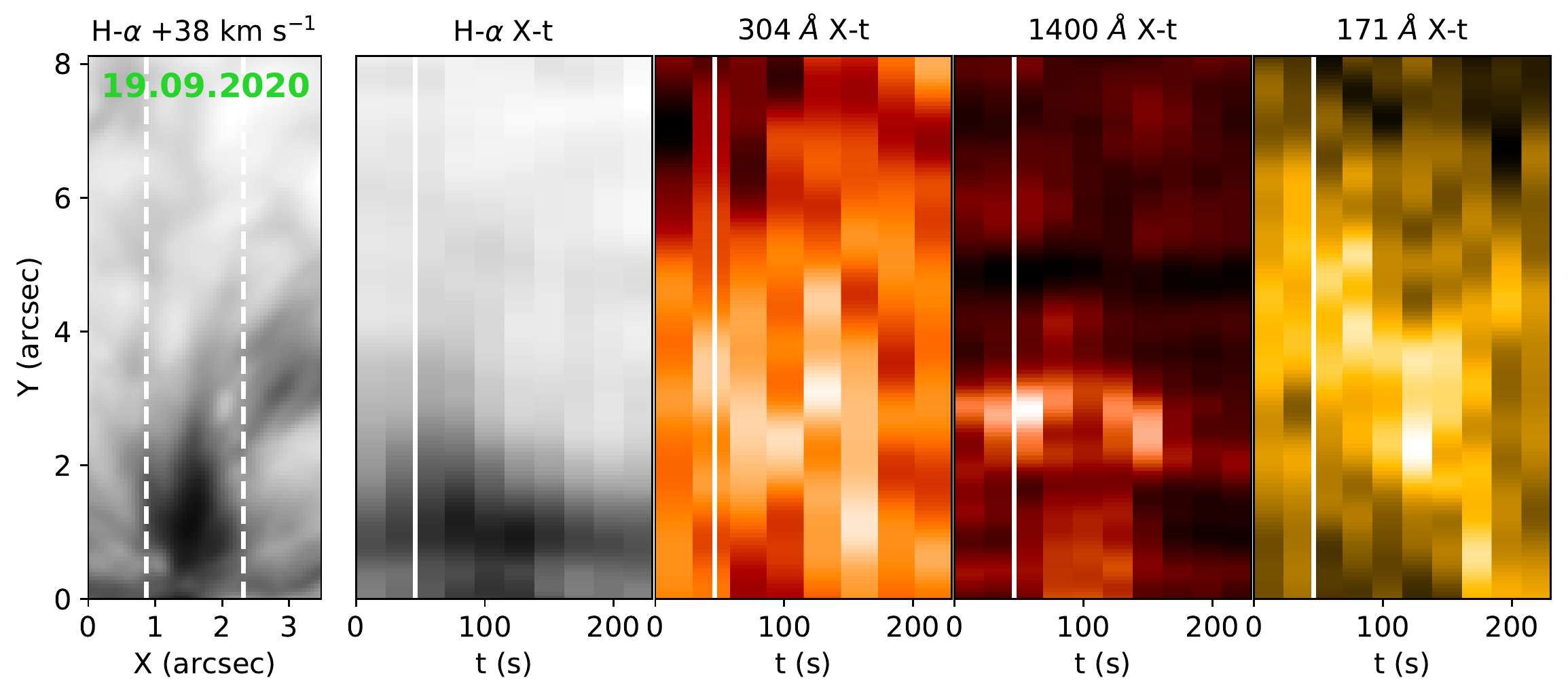}
\caption{\label{figure:DRREs_X-t}%
Space-time ($X$-$t$) diagrams for three representative downflowing RREs on 25 May 2017 (top row) and 19 Sep 2020 (middle and bottom rows) showing their multithermal nature across multiple wavelength channels. The dashed vertical lines in \halpha{} red wing images indicate the region along which the $X$-$t$ maps have been extracted in the coordinated SST, IRIS and SDO datasets. The different panels in each row corresponds to the $X$-$t$ maps in the different wavelength channels as indicated on the top. The solid white vertical lines in the $X$-$t$ maps corresponds to the time step at which these maps have been shown. Animations of the above figures are available at \url{https://www.dropbox.com/sh/qd5s0gko6wcgor7/AACVtF5B9ejVGTfFdscWh8o4a?dl=0}.
}
\end{figure*}
\subsection{Transition region and coronal response to downflowing RREs}
\label{Subsection:TR_coronal_response}
%%-----------First highlight the morphology and time evolution from x-t maps----
%moved figure higher

%%----------Start writing below-----------------------
\begin{figure*}[!ht]
\centering
\includegraphics[bb = 0 10 730 520, width=\textwidth]{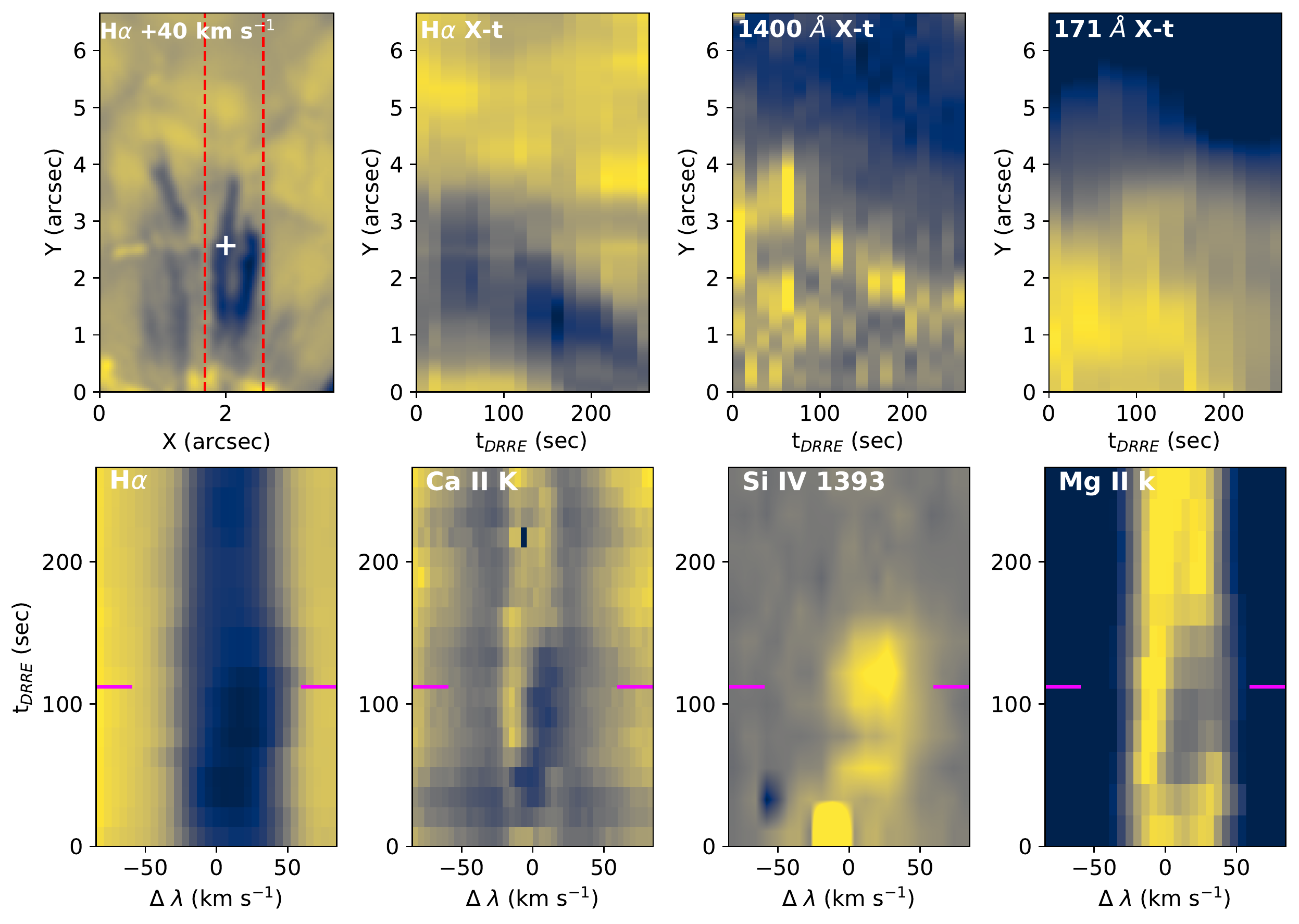}
\includegraphics[bb = 0 10 770 260, width=\textwidth]{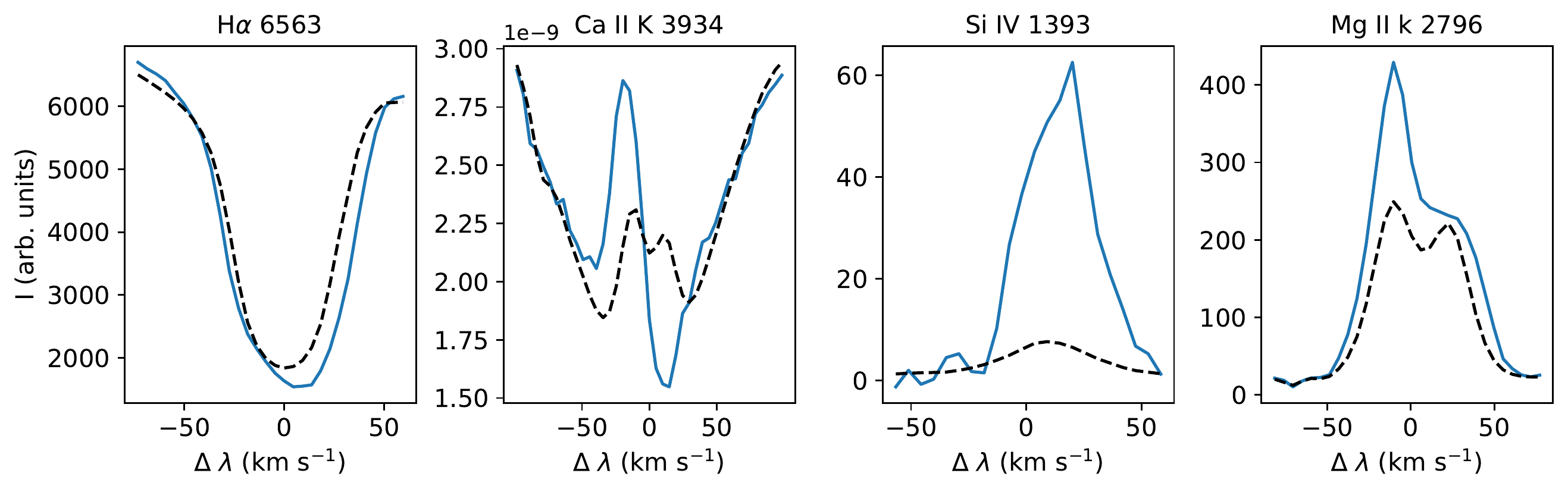}
\caption{\label{figure:RBE_DRRE_Mg_Ca1}%
A representative example of a downflowing RRE observed in SST, IRIS and SDO datasets acquired on 25 May 2017 and its spectral signature across the different chromospheric and transition region spectral lines. Top row (left to right): \halpha{} red wing image showing the downflowing RRE at +40~\kms{}, $X$-$t$ maps (extracted from the region bounded by the two dashed red lines) corresponding to \halpha{} red wing at +40~\kms{}, IRIS \Si{}~1400~\AA\ SJI, and AIA 171~\AA\ channels. Middle row (from left to right): shows the $\lambda t$ diagrams in \halpha{}, \cak{}, \Si{}~1393~\AA\, and \Mgk{}, respectively, corresponding to the spatial location marked with a white cross in the \halpha{} red wing image; while the bottom row shows the corresponding spectral profiles for the different wavelengths at the instant of maximum redward excursion (magenta marker) indicated in the $\lambda t$ diagrams. The dashed spectral profiles corresponds to the spatial and temporal average over the whole time series in each case. The observed intensity is shown in arbitrary units.
}
\end{figure*}

Figure~\ref{figure:DRREs_X-t} shows three examples of downflowing RREs along with their signatures in the transition region and coronal passbands. The vertical dashed lines in the figures show the region which is used to construct the $X$-$t$ maps to study their evolution. The top row of the figure and its associated animation shows the evolution of a typical downflowing RRE observed in dataset~1 in the far red wing (+40~\kms{}) of \halpha{} and its corresponding signature in 
%AIA \ion{He}{II}~304, \ion{Fe}{VIII}~131, and \ion{Fe}{IX}~171~\AA\ passbands. 
the AIA 304~\AA, 131~\AA, and 171~\AA\ passbands. 
%These passbands are dominated by lines from ionized plasma that are formed primarily at 0.1, 0.4 and 0.8~MK, respectively \citep{2012SoPh..275...17L}. 
These passbands are dominated by spectral lines from \ion{He}{II}, \ion{Fe}{VIII}, and \ion{Fe}{IX} ions that are formed primarily around 0.1, 0.4 and 0.8~MK, respectively \citep{2012SoPh..275...17L}. A detailed analysis of the synthetic emission of various AIA passbands from 3D MHD simulations by \citet{2011ApJ...743...23M}, indicated that the above mentioned channels were least influenced by non-dominant ionic species, thereby suggesting that the major contribution to the emission mostly comes from the dominant ions specified earlier.
The apparent inward motion (i.e. towards the bright points) is clearly revealed from the $X$-$t$ maps and also from the animation. While the \halpha{} $X$-$t$ map shows signatures of the downflow starting around $t=170$~s, a close look at the corresponding maps in the hotter passbands of AIA, especially the 171~\AA\ channel, show that the feature is already present close to $t=0$~s. The 304 and 131~\AA\ channels also show faint signatures towards the beginning of the evolution. The animation associated with the \halpha{} red wing image also shows a small horizontal displacement of the bright network element close to the footpoint of the downflowing RRE from \textasciitilde($X$,$Y$)=(2\farcs3,0\farcs5) to \textasciitilde($X$,$Y$)=(1\farcs8,0\farcs5) during its evolution. This transverse displacement can have a LOS component causing the apparent downward motion, but given the fact that the displacement (by \textasciitilde0\farcs5) is small and strictly in the horizontal direction with a velocity of roughly 2~\kms{}, it is unlikely to affect the vertically downward apparent motion shown by the downflowing RRE which has a significantly higher velocity of roughly 30~\kms{} (deduced from the \halpha{} $X$-$t$ map). Moreover, the location of the coronal plasma with respect to the chromospheric downflowing RRE seems to vary substantially with temperature, with a clear spatial offset of the hottest emission above the cooler chromospheric plasma. Since the accuracy of the cross-alignment between the AIA channels is found to be better than the AIA pixel scale, it is likely that this offset (of roughly 1\arcsec) is attributable to the downflowing RRE. This suggests the possibility that the downflowing RREs have a multithermal nature and can be heated to at least lower coronal temperatures roughly at 0.8~MK.

The lack of signature in the first half of the \halpha{} $X$-$t$ map could be primarily because the downflowing RREs, like traditional RREs and RBEs, have a wide range of Doppler shifts associated with them \citep{2016ApJ...824...65P,2021A&A...647A.147B}. Therefore, it is possible that observations in the wing positions, closer to the line core of \halpha{} spectral line, can overcome this gap to some extent. This can be further supported by taking a closer look at the \halpha{} spectral-time evolution diagrams of some additional examples from dataset~1 (shown in Figs.~\ref{figure:RBE_DRRE_Mg_Ca1}, \ref{figure:RBE_DRRE_Mg_Ca2}, and \ref{figure:RBE_DRRE_Mg_Ca3} and described further in Sect.~\ref{Subsection:Spectral_signature}), which provides an impression that the downflowing RREs have roughly zero Doppler shifts during their initial phases which increases gradually with time.

% The lack of signature in the first half of the \halpha{} $X$-$t$ map could be attributed to two main reasons: Firstly, downflowing RREs, like traditional RREs and RBEs, have a wide range of Doppler shifts associated with them \citep{2016ApJ...824...65P,2021A&A...647A.147B}. Therefore, it is possible that observations in the wing positions, closer to the line core of \halpha{} spectral line, can overcome this gap to some extent. Secondly, the presence of a strong emission in the 171~\AA\ channel could be an indication that during the initial stages, the downflowing RRE is too hot to be detected in \halpha{} observations because neutral hydrogen will most likely be ionized. 

The middle and the bottom rows of Fig.~\ref{figure:DRREs_X-t} show two additional examples of multithermal downflowing RREs from dataset~2 in the same format described above. Like before, here also we find emission signatures in the transition region and coronal passbands corresponding to the chromospheric downflows. The IRIS \Si{}~1400~\AA\ channel, sampling roughly 0.08~MK, is shown in these examples instead of the AIA 131~\AA\ passband, since the signatures in the latter is not as prominent. The $X$-$t$ maps appear to be coarser in comparison to the earlier example because the temporal cadence of dataset~2 (23.6~s) is lower than dataset~1, and the downflows last for a shorter period of time. The animation associated with this figure provides a clearer impression of the apparent motions associated with the two events. Unlike the example in the top row, here we see a cotemporal downflow in all channels including the \halpha{} red wing. This could be due to the image being observed at a line position that is closer to the \halpha{} core. Based on the \Si{}~1400~\AA\ $X$-$t$ maps, we also find that the plasma associated with the downflowing RREs is at least sometimes heated to transition temperatures throughout its evolution. It is also to be noted that, unlike the middle row, the \Si{} channel corresponding to the downflowing RRE in the bottom row does not show significant emission towards the later stages of evolution starting roughly around 150~s. The hotter 171~\AA\ channel also shows strong emission throughout the evolution of the downflowing RREs except in the middle row where the signal starts to appear only towards the later half. Similarly, the 304~\AA\ passband shows a gradual enhancement in emission as the evolution progresses with a substantial emission around $t=100$~s (middle row). The spatial offset between the hotter coronal and cooler chromospheric plasma is even more prominent in the bottom row indicating that the hotter material lies on top of the cooler plasma. The scenario described here suggests that downflows are not completely heated to transition region (or even coronal temperatures) and are likely multi-threaded with cold and hot threads. We also note that the downflowing RRE shown in the middle row has a smaller spatial extent in comparison to the other two rows, roughly covering 6--7 AIA pixels in the $Y$ direction. However, the apparent motion in the AIA channels show strong emission that is well correlated with the corresponding signature in the \halpha{} $X$-$t$ map. This suggests that some downflows can also have a small blob-like appearance, contrary to elliptical shapes that is commonly observed in spicules.

In both examples, the $X$-$t$ maps suggest that the apparent velocity of the chromospheric, transition region and coronal plasma is probably very similar, with velocities in the range between 25--40~\kms{}, and they move (inward) towards the bright network regions. It is to be noted that we did not find any associated signature in the blue-wings of \halpha{} preceding the downflowing RREs. This has already been shown to be the case for a few downflowing RREs by \citet{2021A&A...647A.147B}, and will be discussed in detail in the following section. Moreover, we also suggest an alternate mechanism from the numerical simulations in Sect.~\ref{subsection:discussion_loop_drre}, that could likely explain the appearance of such downflows without preceding blueward excursions.

\subsection{Spectral signatures}
\label{Subsection:Spectral_signature}

\begin{figure*}[!ht]
\centering
\includegraphics[bb = 0 10 730 520, width=\textwidth]{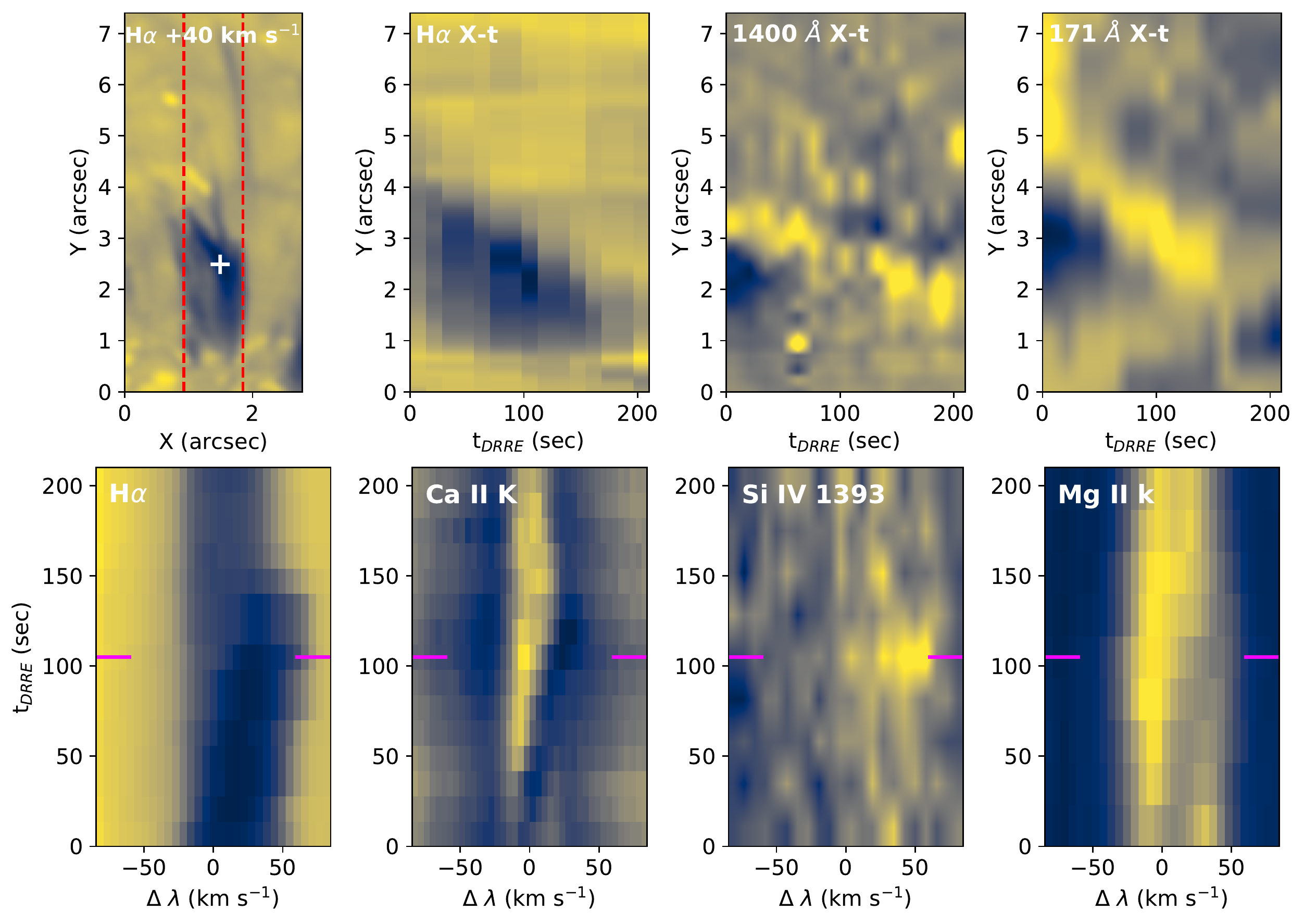}
\includegraphics[bb = 0 10 770 260, width=\textwidth]{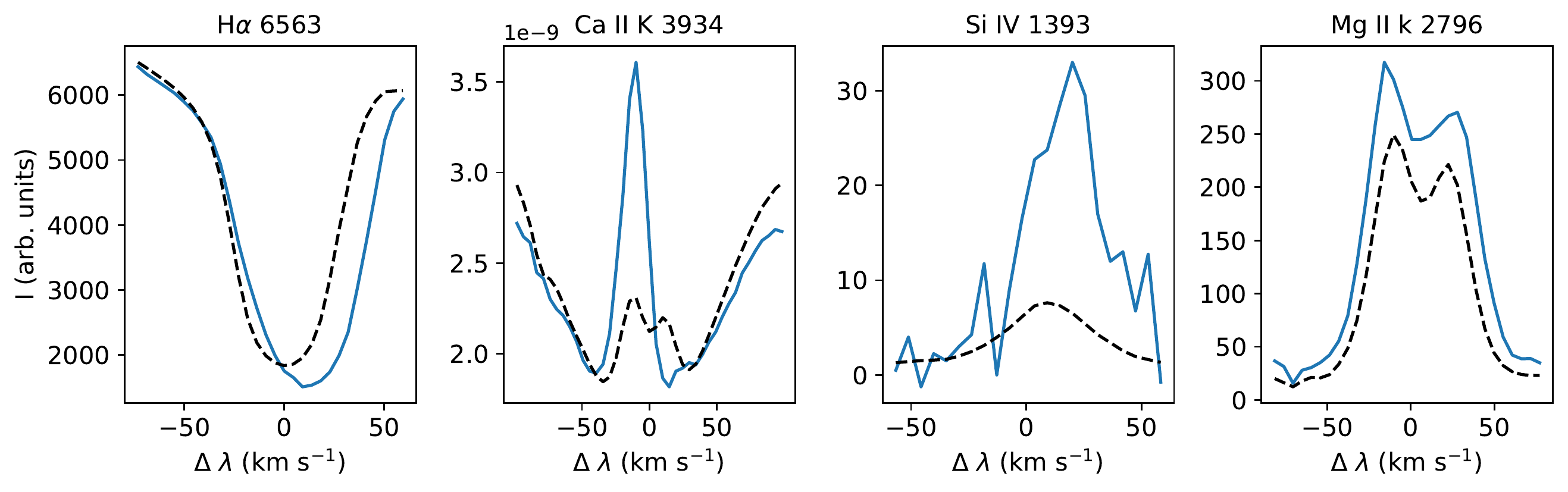}
\caption{\label{figure:RBE_DRRE_Mg_Ca2}%
Second representative example of a downflowing RRE observed on 25 May 2017 in the same format as Fig.~\ref{figure:RBE_DRRE_Mg_Ca1}.
}
\end{figure*}

\begin{figure*}[!ht]
\centering
\includegraphics[bb = 0 10 730 520, width=\textwidth]{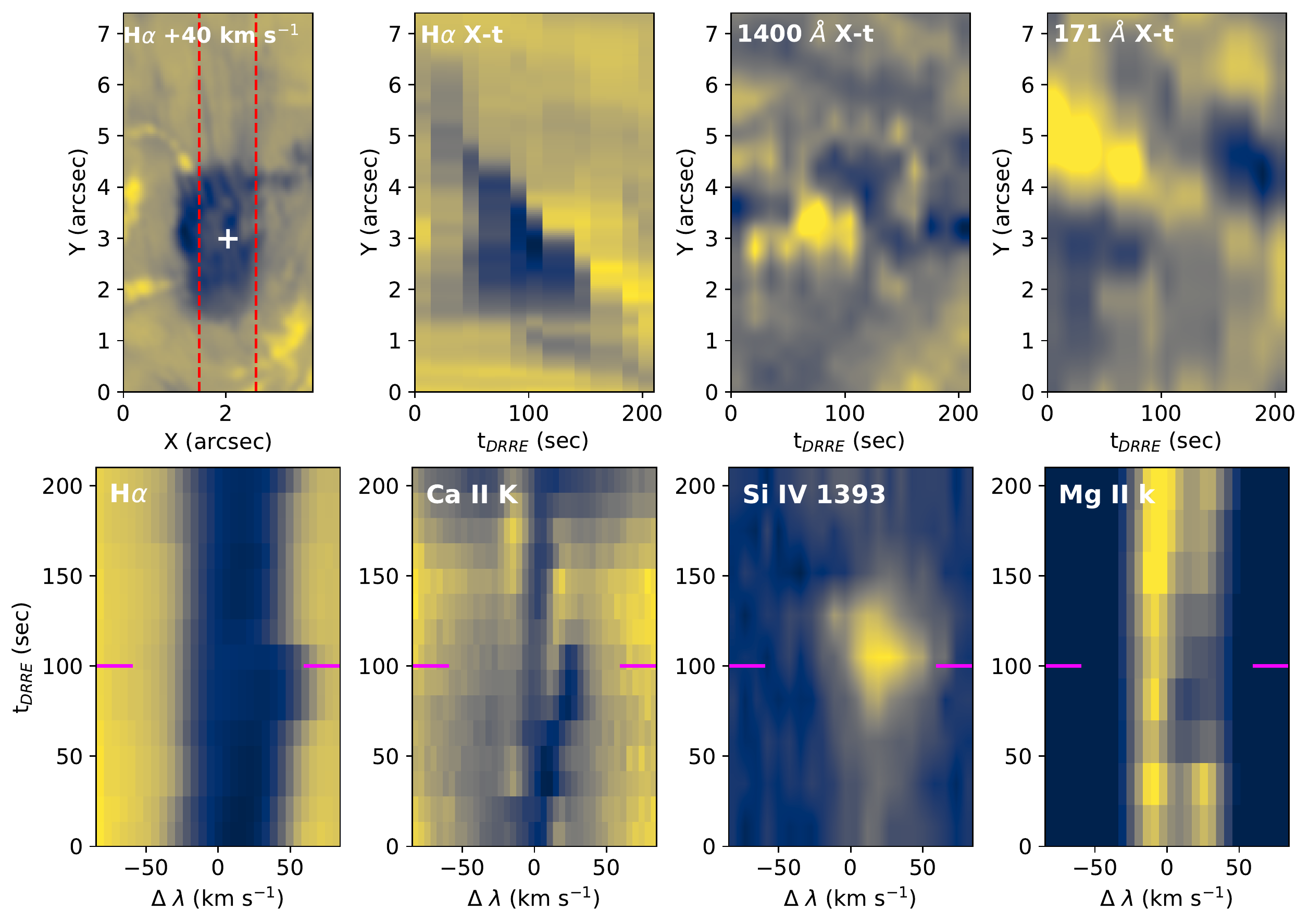}
\includegraphics[bb = 0 10 770 260, width=\textwidth]{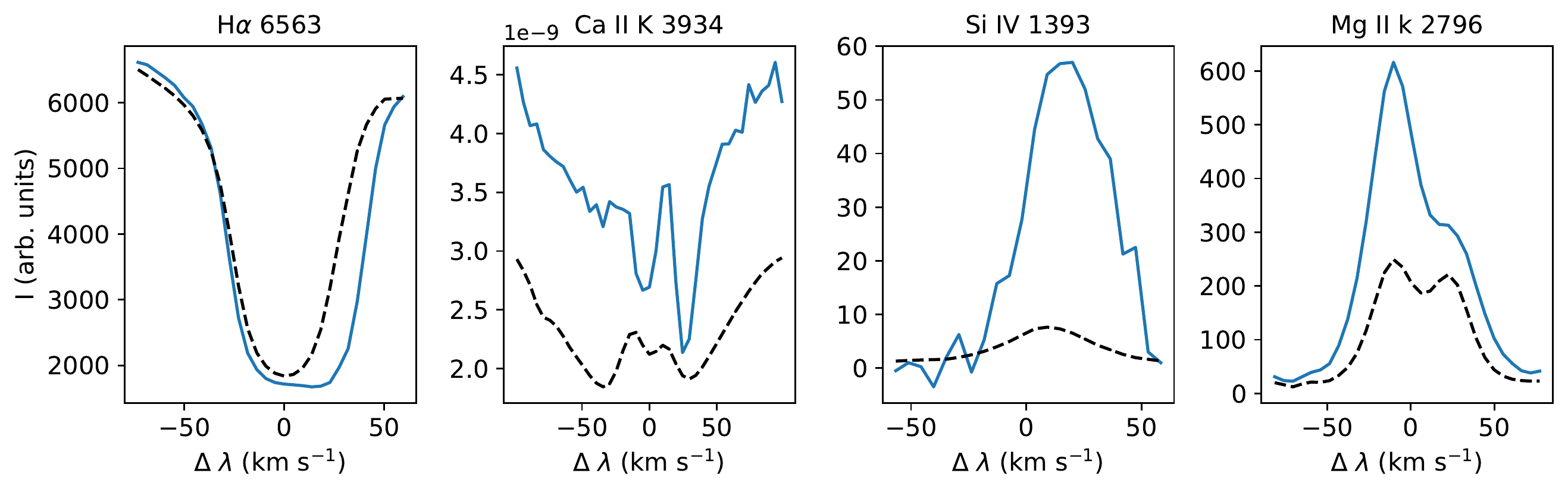}
\caption{\label{figure:RBE_DRRE_Mg_Ca3}%
Third representative example of a downflowing RRE observed on 25 May 2017 in the same format as Fig.~\ref{figure:RBE_DRRE_Mg_Ca1}.
}
\end{figure*}
%---------Write below---------------
The spectral signatures of the downflowing RREs in the solar chromosphere (mainly in \halpha{}) have already been discussed earlier in \citet{2021A&A...647A.147B}. In this paper, we aim to show their signatures observed in the lines formed in the upper chromosphere and transition region. 
%Unlike, SST observations, we are limited by the spatial coverage of the IRIS rasters in the FOV. 
The spatial coverage of the IRIS rasters is limited as compared to the SST FOV. 
Therefore, the number of downflowing RREs overlapping on the IRIS raster at a given time is limited. Moreover, despite being quasi-simultaneous, there could be residual temporal offsets between the SST and IRIS rasters of the order of 3--4~s which can cause further limitation. The examples described in Sect.~\ref{Subsection:TR_coronal_response} do not overlap the IRIS rasters and therefore have no spectral information. In this section, we highlight a few examples from dataset~1 where the downflowing spicular events overlapped co-temporally and co-spatially in all three (SST, IRIS--SJI+raster, and SDO) instruments.

Figure~\ref{figure:RBE_DRRE_Mg_Ca1} shows an example of a downflowing RRE with the temporal evolution indicated with the help of $X$-$t$ maps in the different passbands in the top row, while the middle row shows the spectral evolution in the form of $\lambda t$ (spectral--time diagrams) in \halpha{}, \cak{}, \Si{}, and \Mgk{}, respectively from left to right. The dashed red lines in the \halpha{} red wing image at +40~\kms{} indicates the region selected for generating the $X$-$t$ maps for the different passbands, which clearly show the apparent inward motion in all the channels. This suggests enhancement to temperatures as high as 0.8~MK during the downflow based on the emission pattern in 171~\AA. Furthermore, a closer look at the $X$-$t$ diagrams reveal that \halpha{} shows fainter signal in the period between 0 and 100~s whereas the \Si{}~1400~\AA\ and 171~\AA\ channel shows significant emission. This could hint that during the initial stages, the downflow is too hot to be detected clearly in the \halpha{} red wing images, which was perhaps also the case in the example shown in the top row of Fig.~\ref{figure:DRREs_X-t}. Such behavior is compatible with a multithermal nature in the downflowing RREs.

The $\lambda t$ diagrams reveal the temporal evolution of the spectral profiles at a given spatial location of the downflowing RRE. The excursion towards the redward side of the \halpha{} line center is nearly co-temporal with the excursions of the central absorption K$_{3}$ (k$_{3}$) feature in the \cak{} (\Mgk{}) diagrams, and a strong redward emission in the \Si{}~1393 diagram. We see a typical development of a highly asymmetric line profile peaking out at 20--40~\kms{} redward of the nominal line center in all the cases. This is further illustrated in the bottom row of Fig.~\ref{figure:RBE_DRRE_Mg_Ca1}, where spectral profiles corresponding to the time of maximum redward excursion in \halpha{} are shown. The \cak{} and \Mgk{} spectral lines show a red shift in the central absorption K$_{3}$ (k$_{3}$) feature with respect to the average profiles (constructed from temporal and spatial average over the entire FOV for \cak{} and over all the slit positions for \Mgk{}), which causes the K$_{2r}$ and k$_{2r}$ emission peaks to disappear completely. Furthermore, an enhancement in the K$_{2v}$ and k$_{2v}$ peaks is also seen that makes it well aligned with the observations by \citet{Luc_2015} and optically thick line formation mechanism in spicules described in \citet{my_paper_3}.

The \Si{}~1393~\AA\ spectra are very noisy at this exposure time (2~s). Nevertheless, a comparison of the $\lambda t$ diagram with other panels show similar temporal behavior with redward asymmetric excursions that are in tandem with their pure chromospheric counterparts \halpha{}, \cak{} and \Mgk{}. Preliminary analysis based on single Gaussian fits to the far-UV spectra indicate a Dopplershift of the order of 15--20~\kms{} for the downflowing RRE which is slightly higher than the values reported by \cite{Luc_2015} for traditional RREs.

Figures~\ref{figure:RBE_DRRE_Mg_Ca2} and \ref{figure:RBE_DRRE_Mg_Ca3} show two additional examples of downflowing RREs in the same format as Fig.~\ref{figure:RBE_DRRE_Mg_Ca1}. Like the earlier cases, we clearly see the multithermal behavior of the downflowing RREs based on their near co-temporal visibility in the 1400~\AA\ and 171~\AA\ along with \halpha{} red wing $X$-$t$ maps. A noticeable distinction among the three examples is that the emission pattern of 171~\AA\ shown in Fig.~\ref{figure:RBE_DRRE_Mg_Ca1}, appears to occur over a larger spatial region and is co-located with the appearance of the dark spicular region in \halpha{} in comparison to the other cases, where we clearly see the emission confined to narrower bright streaks. This difference is also apparent to some extent in the examples shown in Fig.~\ref{figure:DRREs_X-t}. Furthermore, the 171~\AA\ emission seems to weaken towards the later part of the evolution of the downflowing RREs (in Figs.~\ref{figure:RBE_DRRE_Mg_Ca2} and \ref{figure:RBE_DRRE_Mg_Ca3}), though the inward apparent progression remains evident. Contrary to Figs.~\ref{figure:RBE_DRRE_Mg_Ca1} and  \ref{figure:RBE_DRRE_Mg_Ca2}, the emission pattern in the 1400~\AA\ $X$-$t$ map observed in Fig.~\ref{figure:RBE_DRRE_Mg_Ca3}, lasts for a shorter period of time (\textasciitilde30~s) around $t=100~s$. The inward apparent motion (downflow) is also not very prominent like in the other wavelength channels. This could be mainly attributed to either the low exposure time of the acquired SJIs or the downflowing RRE not being hot enough throughout its evolution or a combination of the two. The evolution in the adjacent 171~\AA\ map seems to support the scenario that the downflowing RRE indeed cools down after $t=100~s$ (or so), since the emission is significantly reduced. Also, with the exception of Fig.~\ref{figure:RBE_DRRE_Mg_Ca1}, at any given time the emission corresponding to 171~\AA\ appears to have a slight offset in the $Y$ direction in comparison to the cooler chromospheric component. This seems to also be the case when the transition region brightenings are compared with \halpha{}. This is similar to the examples shown in the top and bottom rows of Fig.~\ref{figure:DRREs_X-t}. Therefore, it suggests the possibility of a scenario where the hot plasma (\textasciitilde0.1--0.8~MK) lies above the cooler chromospheric component at \textasciitilde0.01~MK, once again revealing the multithermal nature of downflowing RREs, similar to RBEs and RREs \citep{2011Sci...331...55D,Tiago_2014_heat,2014ApJ...795L..23S}.

The $\lambda t$ diagrams and the corresponding spectral profiles at the time of maximum excursion of the downflowing RREs show similar properties to what has been described earlier in Fig.~\ref{figure:RBE_DRRE_Mg_Ca1}. In fact, the two downflowing RREs shown in Figs.~\ref{figure:RBE_DRRE_Mg_Ca2} and \ref{figure:RBE_DRRE_Mg_Ca3} show stronger absorption asymmetries towards the redward side of the line center both in \halpha{} and \cak{}. Even the emission corresponding to \Si{}~1393~\AA\ appears to have a stronger excursion when compared against Fig.~\ref{figure:RBE_DRRE_Mg_Ca1}. A closer look at the \cak{} spectral line shown in the bottom row of Fig.~\ref{figure:RBE_DRRE_Mg_Ca3} reveals significant enhancement in its specific intensity (including the line wings) in comparison to the \cak{} spectral profiles shown in the other examples. This is attributed to the bright network regions in the background of the spicule which is clearly visible in the \halpha{} red wing image. Nevertheless, the asymmetry in the line profile with a shift of the K$_{3}$ line core towards the redward side of the line center, along with the enhancement in the K$_{2v}$ peaks is apparent in all the three cases. The red shift of the k$_{3}$ core in the \Mgk{} spectral profile shown in the bottom row of Fig.~\ref{figure:RBE_DRRE_Mg_Ca2}, does not completely eliminate or suppress the k$_{2r}$ peak as evident in the other examples. However, like \cak{}, the asymmetry is clear enough with the k$_{2v}$ peak being enhanced in comparison to the k$_{2r}$ indicating the presence of strong velocity gradients in spicules that remove the upper-layer opacities at the k$_{2r}$ wavelengths. 

In the examples described in this section, the $\lambda t$ diagrams (for all wavelengths) do not show signs of preceding blue shift that are typical for type-I spicules. This is suggestive of an alternate mechanism (based on loops), that could also be responsible for the observed downflowing RREs. This is discussed further in Sect.~\ref{subsection:discussion_loop_drre}. We also find that the Dopplershifts associated with the downflowing RREs (25--50~\kms{}) are compatible with their apparent velocities of the corresponding transition region and coronal brightenings. The latter indicates that the downflows studied here are basically a result of real plasma (mass) flows that are different than the high-speed (100--300~\kms{}) apparent motions often associated with transition region network jets \citep{2014Sci...346A.315T}. Such high-speeds in the network jets are possibly caused by rapidly propagating heating fronts rather than mass flows \citep{2017ApJ...849L...7D}.

The properties of the \halpha{}, \cak{}, and \Mgk{} line profiles for the downflowing RREs described in this section are well aligned with the spectral characteristics of RREs/downflowing RREs reported by \citet{my_paper_3}. Characterizing millions of RBE and RRE spectral profiles belonging to \cak{}, \halpha{}, and \Mgk{} with the statistical $k$-means clustering technique, they found that opacity shifts with strong gradients in the LOS velocity are dominant among spicules. Interestingly, they also reported that the RRE/downflowing RRE-like \cak{} and \Mgk{} spectral profiles are, on average, broader than the RBE-like profiles, and have significantly enhanced K$_{2v}$ (k$_{2v}$) peaks when compared both against the average profiles (over the whole FOV and time series) and RBE-like profiles \citep[refer to Fig.~3 in][]{my_paper_3}. This hinted a possibility of other mechanisms, likely heating-based, in the RREs/downflowing RREs. The chromospheric spectral characteristics shown in this study clearly correlate with the earlier observations and point towards a possibility where the downflowing RREs do undergo heating during their evolution. In the following section, we investigate this scenario with the help of an advanced MHD simulation.

\subsection{Comparing synthetic observables from MHD simulation}
\label{Subsection:synthetic_numeric}

\begin{figure*}
   \centering
   \includegraphics[width=\textwidth]{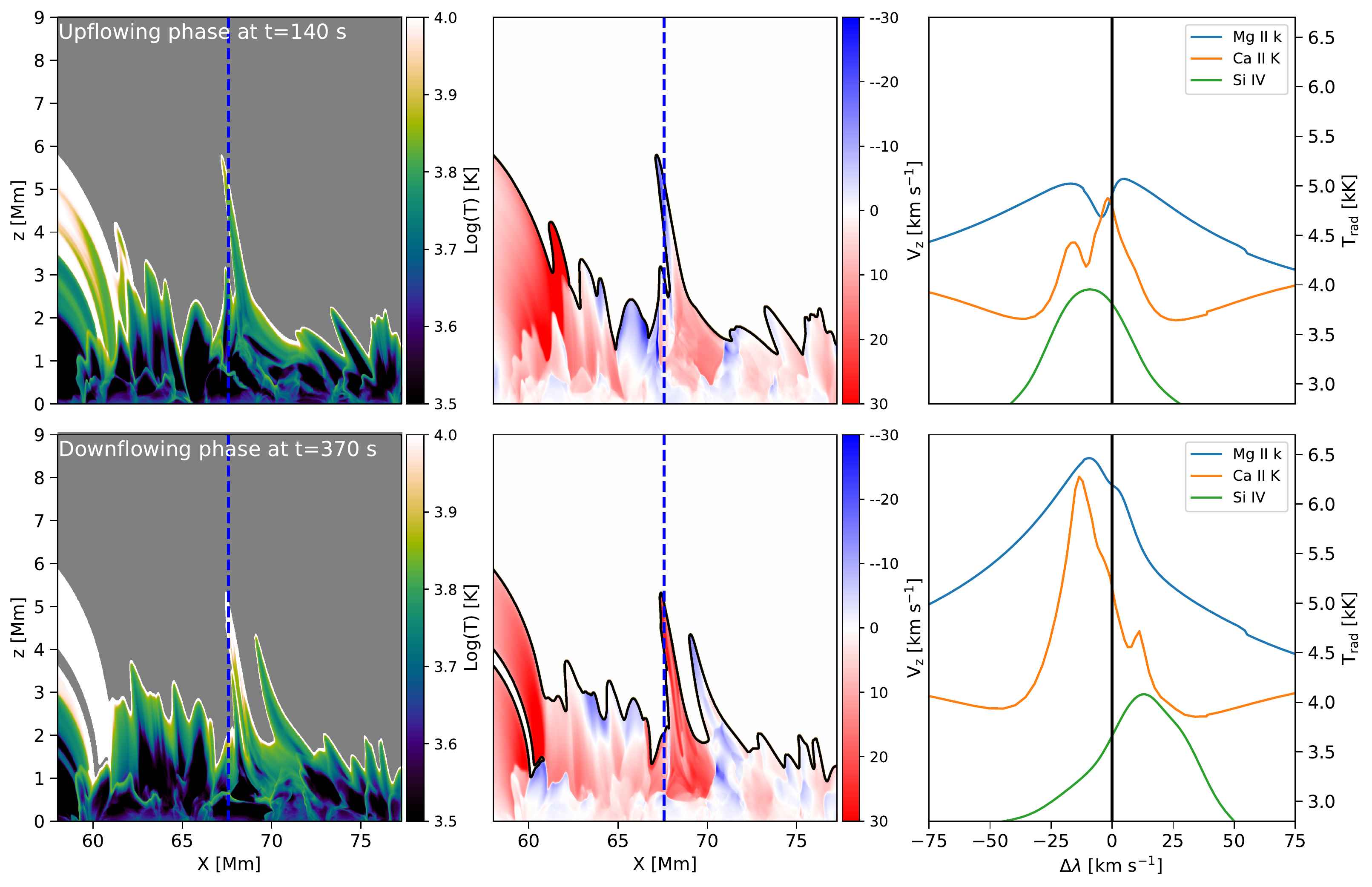}
   \caption{Synthetic observables of the chromospheric and transition region spectral lines obtained from the numerical simulation. Top row (left to right) shows a cutout of the 2D temperature map, the signed vertical velocity around spicule~1, and \cak{}, \Mgk{}, and \Si{}~1393~\AA\ synthetic spectra corresponding to the upflowing phase of the spicule at t=140~s. Bottom row shows the corresponding temperature, vertical velocity, and synthetic spectra for the downflowing phase of the spicule at t=370~s in the same format as the top row. Temperatures beyond 30~kK have been masked in the 2D temperature maps and is indicated with a black contour in the vertical velocity maps. The dashed vertical blue line corresponds to the LOS of the observer along which optically thick/thin radiative transfer is performed.
   } 
        \label{figure:synthetic_RBE_DRRE_Mg_Ca}%
    \end{figure*}

%%------------------------------------------------------

\subsubsection{Radiative transfer diagnostics}
\label{Subsubsection:Rad_transfer_diagnostics}

A valid comparison between any numerical simulation and observation is only possible by accounting for the radiation transport of photons through the simulated stellar atmosphere, since the only source of information available from observations is the intensity at specific wavelengths of interest. In this section, we show synthetic spectra for a typical spicule (spicule~1 in Fig.~\ref{figure:Context_v1_sim}) for the lines formed in chromosphere and transition region. The details of the simulation have been described in Sect.~\ref{subsection:numerical_sim}.

We tracked the complete evolution of spicule~1 but we restrict ourselves to the two phases shown in Fig.~\ref{figure:synthetic_RBE_DRRE_Mg_Ca} for the sake of brevity. Phase~1 in the top row is the rising or upflowing phase of the spicule shown at roughly 140~s after the start of the simulation, while phase~2 (bottom row) is the downflowing phase which occurs about 230~s later. The animation associated with Fig.~\ref{figure:Context_v1_sim} shows that the upflowing phase is characterized by strong initial velocities of the order of 50--70~\kms{} towards the observer (the direction is relative to the observer looking vertically downwards), while the downflowing phase has strong velocities of at least 30~\kms{} away from the observer. These velocities are similar to what has been observed in the on-disk manifestations of type-II spicules such as RBEs, and RREs/downflowing RREs \citep{Luc_2009,Sekse_2012,2013ApJ...769...44S,Vasco_2016,2021A&A...647A.147B}. In the simulation, we refer to phase~2 or the returning phase of spicule~1 as a downflowing RRE.  

%The last column of Fig.~\ref{figure:synthetic_RBE_DRRE_Mg_Ca} shows the synthetic \cak{}, \Mgk{} and \Si{}~1393~\AA\ spectra corresponding to the dashed blue vertical marker displayed in the temperature and velocity maps, for both the upflowing and downflowing phases. 
In the last column of Fig.~\ref{figure:synthetic_RBE_DRRE_Mg_Ca} synthetic \cak{}, \Mgk{} and \Si{}~1393~\AA\ spectral profiles for both the upflowing and downflowing phases are shown.
Radiative transfer computations were performed along the direction of the dashed blue vertical marker with the observer looking down. The upflowing phase in the optically thick lines show distinct shifts in the K$_{3}$ (k$_{3}$) core towards the blueward side of the respective line centers that suppresses the K$_{2v}$ (k$_{2v}$) peaks, while causing an enhancement in the opposite emission peaks. The trend is quite pronounced for the \cak{} spectral line but to a lesser extent in \Mgk{}. This is because the two spectral lines are sensitive to slightly different atmospheric regimes which affects their line formation mechanism. This will be discussed further in Sect.~\ref{Subsubsection:Cont_func_analysis}. The hotter \Si{} profile shows a blueward shift of its peak that is well correlated with K$_{3}$ and k$_{3}$ line cores. 

The \cak{} and \Mgk{} profiles corresponding to the downflowing stage at the same spatial location show significant enhancement in their radiation temperatures in comparison to the upflow. Furthermore, the shift of the K$_{3}$ (k$_{3}$) core towards the redward side of the respective line centers is so strong, that it completely makes the K$_{2r}$ (k$_{2r}$) peak disappear. The local emission peak corresponding to the orange \cak{} profile at $\Delta \lambda \approx 15$~\kms\ can be mistakenly interpreted to be the reduced K$_{2r}$ (and the observed "valley" immediately blueward of this peak to be K$_{3}$), but a detailed analysis based on contribution function suggests that it is not the case. We refer the reader to Sect.~\ref{Subsubsection:Cont_func_analysis} for a more in-depth discussion. For the \Si{} emission line, not only is the peak strongly red shifted, the line also appears to be broader compared to its upflowing counterpart. The radiation temperature, corresponding to the peak emission, is also slightly higher than the upflowing phase but overall the \Si{} line has weaker emission in comparison with the optically thick lines. These observations are well correlated with the spectral signatures of the downflowing RREs described in Sect.~\ref{Subsection:Spectral_signature} and Figs.~2 and 3 of \citet{my_paper_3}. Moreover, a close comparison of the temperature maps shown in the first column of Fig.~\ref{figure:synthetic_RBE_DRRE_Mg_Ca} reveal a marked increase in temperature towards the tip of the spicule during the downflowing phase. This increase in temperature could indicate signs of heating during the downflowing phase of spicules. This forms the basis of the results presented in Sect.~\ref{Subsubsection:heating_simulations}.

\subsubsection{Heating in the downflowing RREs}
\label{Subsubsection:heating_simulations}

\begin{figure*}[ht] 
  \begin{subfigure}[b]{0.50\textwidth}
    \centering
    \includegraphics[width=0.99\linewidth]{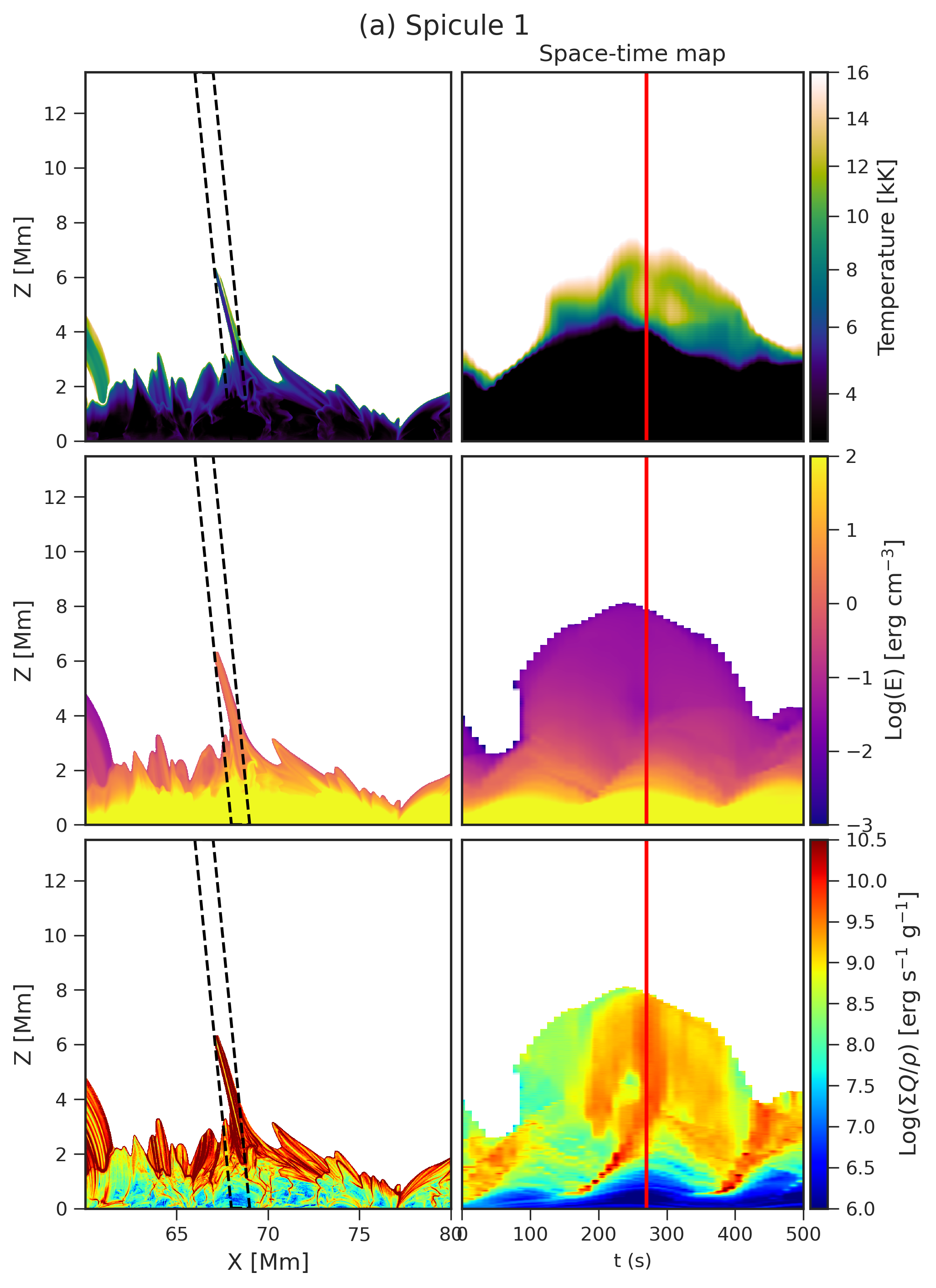} 
    %\caption{Spatio-temporal evolution of spicule~1.} 
    \label{figure:cont_Ca_RBE_2} 
    %\vspace{4ex}
  \end{subfigure}%% 
  \begin{subfigure}[b]{0.50\textwidth}
    \centering
    \includegraphics[width=0.99\linewidth]{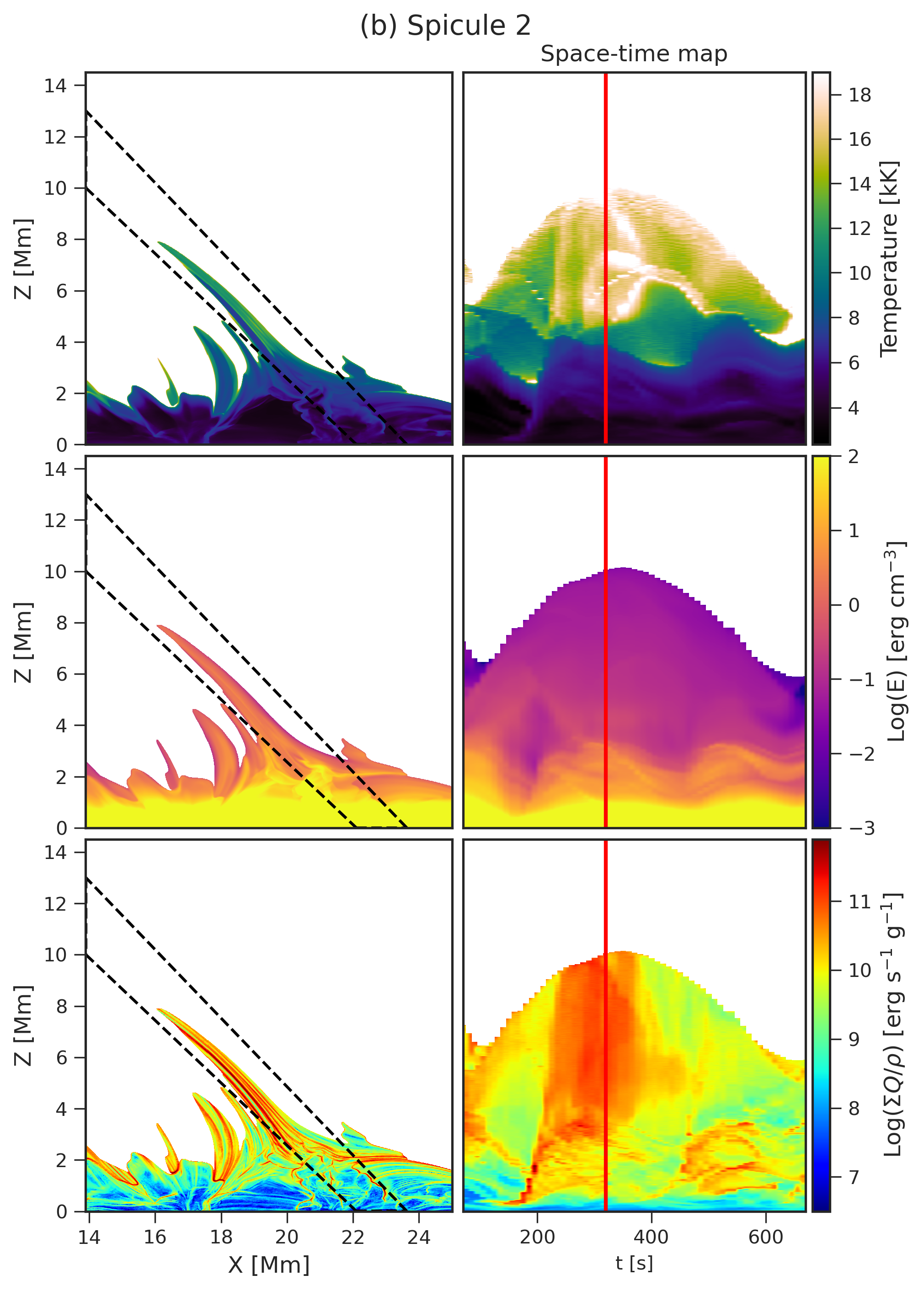} 
    %\caption{Spatio-temporal evolution of spicule~2.} 
    \label{figure:cont_Ca_DRRE_2} 
   % \vspace{4ex}
  \end{subfigure} 
  \caption{Spatio-temporal evolution of the different thermodynamic parameters obtained from the MHD simulation. The left column of subfigure~a shows the 2D maps of temperature, internal energy and sum over heating parameters (as described in the text) per particle ($\rho$ being the plasma density), and the right column indicates the space-time maps corresponding to each of the physical quantities for spicule~1. The two columns of subfigure~b shows the spatio-temporal evolution corresponding to the same physical quantities for spicule~2, in the same format as subfigure~a. The dashed lines in the left column of the two subfigures show the bounding regions that are considered for extracting the $X$-$t$ maps (shown in the respective right columns), and the vertical red line in the $X$-$t$ maps corresponds to the time steps at which these maps have been shown. Animations of these subfigures are available online \url{https://www.dropbox.com/sh/zqwp6ipw12n80y8/AADJ7ocF0TraGS8ygBNFLE2-a?dl=0}}
  \label{figure:heating_per_particle} 
\end{figure*}
%%------------------Write below this line-------------------

The analysis presented in Fig.~\ref{figure:synthetic_RBE_DRRE_Mg_Ca} indicate that the downflowing phase of spicule~1 is marked with an enhancement in both the gas (plasma) temperature and radiation temperature of spectral profiles. In this section, we investigate the cause of this enhancement by exploiting the physical parameters obtained from the numerical simulation.

%Subfigures~a and b in Fig.~\ref{figure:heating_per_particle} show the complete thermodynamic evolution of the two spicules of interest.
Figure~\ref{figure:heating_per_particle} shows the complete thermodynamic evolution of the two spicules of interest.
The spicules have different inclination with respect to the LOS of the observer and each subfigure displays two columns, where the left column shows the thermodynamic quantities being investigated (i.e. temperature, energy, and the summation over various heating parameters normalized by the number density) and their corresponding time evolution are indicated with $X$-$t$ maps shown on the right. In both examples, we masked out regions from the FOV where the temperature exceeded 30~kK because both \ion{Mg}{II} and \ion{Ca}{II} ionizes well below this temperature. We consider heating in the spicules due to Joule dissipation, ambipolar diffusion, viscosity, thermal conduction, and radiative terms that can be directly obtained from the Bifrost code. Ambipolar diffusion refers to the diffusion between charged species and neutral particles that plays a key role in the solar atmosphere where the plasma is partially ionized. The radiative terms refer to optically thin losses, radiative losses by neutral hydrogen, singly ionized calcium, and magnesium, and heating by incident UV radiation from the solar corona \citep{2012A&A...539A..39C}.   

The two subfigures and their associated animations show the complete evolution of the two spicules. Upon closer investigation, we see that the temperature starts to show enhancement right at the time when the spicules reach their maximum extent (or completing the upflowing phase), and initiate the downflow. The enhancement is more significant for spicule~2 with temperatures reaching as high as 20~kK during the maximum excursion phase. Overall, in both cases, we find that the temperature increases from 7--9~kK during the start of the upflowing phase and continues to increase until it reaches the peak of the parabolic trajectory, where it reaches as high as 18--20~kK. This is found to be in agreement with the analysis by \citet{Juan_2017_Science,Juan_2018} and \citet{2017ApJ...849L...7D}. The energy maps follow the temperature very closely showing similar parabolic trajectories, but unlike temperature, they do not show any substantial difference between the two phases. This is, in part, related to the fact that the density of the spicules decreases as they evolve. We refer the reader to Sect.~\ref{Subsection:simulation_discussion} for further discussions.

The sum over the heating terms normalized by number density, shown in the last row of the two subfigures in Fig.~\ref{figure:heating_per_particle}, display a compelling visual correlation with the corresponding temperature $X$-$t$ maps in the top row. The correlation is quite striking for spicule~1 where we see that the heating actually increases right around the time when the temperature starts to show enhancement %(\textasciitilde$t=250~s$),
(at about $t=250$~s), and it continues to remain significantly heated covering almost the entire downflowing phase of the spicule. The enhancement in heating is also quite significant for spicule~2, and it also occurs around the same time as the temperature gets enhanced (at about $t=250$~s). However, unlike spicule~1, the heating tends to get weaker towards the end of the downflowing phase. Nonetheless, both spicules clearly show heating towards the end of the upflowing phase and beginning of the downflowing phase, that correlates well with the behavior observed in the temperature maps. Moreover, the temporal duration of enhanced heating is similar in both cases lasting roughly 200~s. 

The results from the MHD simulation presented in this section provide a clear indication that the downflowing phase (downflowing RRE) of the spicules is indeed hotter in comparison to the upflow (RBE) as we inferred from the observed and synthetic spectral profiles.

%In other words, the downflowing phase is hotter and The heating during the evolution of the spicules result from a 

\subsubsection{Contribution function analysis for optically thick lines}
\label{Subsubsection:Cont_func_analysis}

\begin{figure*}[htbp!] 
\centering
  \begin{subfigure}[b]{0.49\textwidth}
   % \centering
    \includegraphics[width=0.99\linewidth,clip=true,trim=0.0cm 0.0cm 0.8cm 0.0cm]{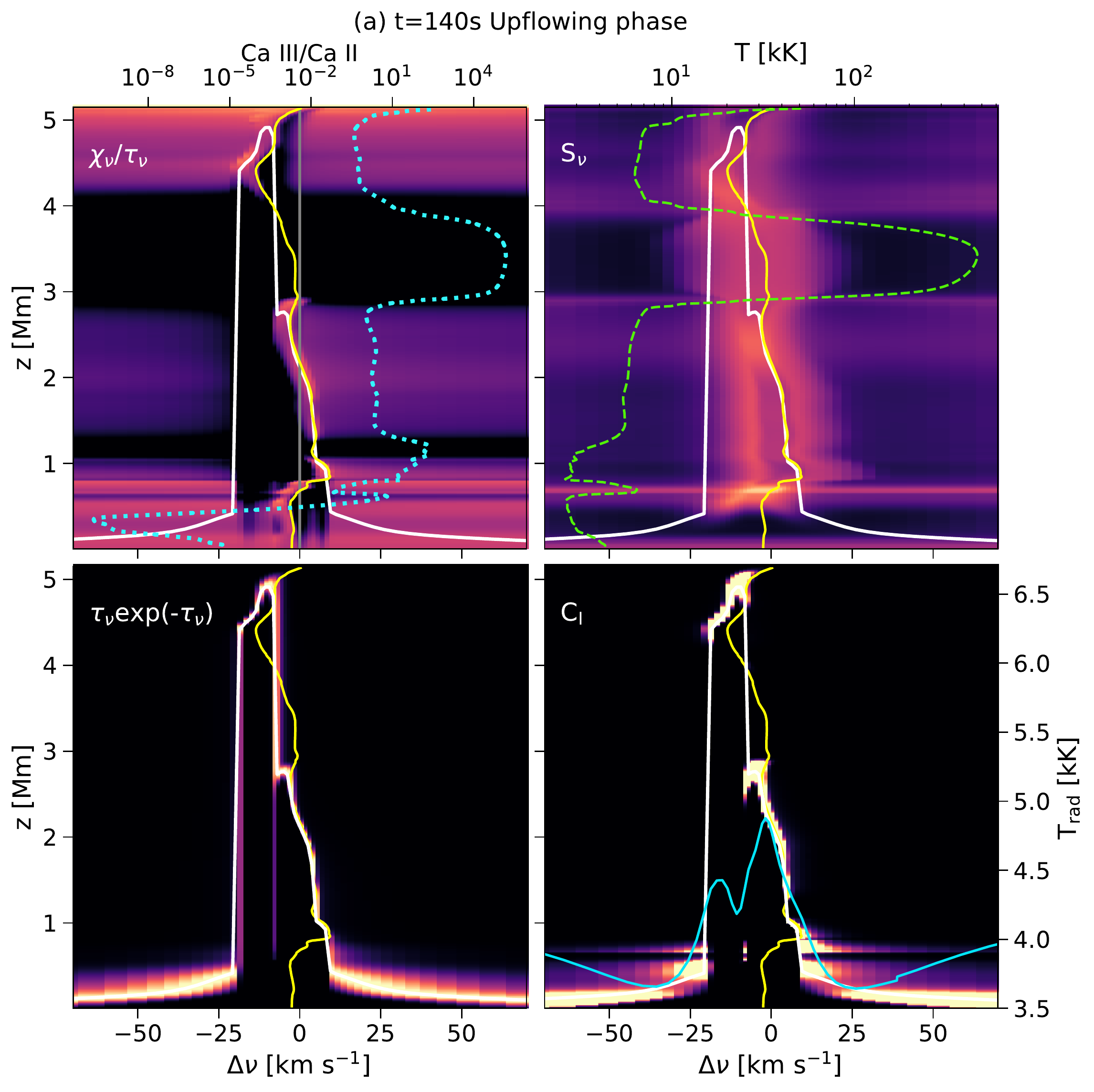} 
 %   \caption{Formation of \cak{} for the upflowing phase.} 
    \label{figure:cont_Ca_RBE} 
  %  \vspace{4ex}
  \end{subfigure}%% 
  \begin{subfigure}[b]{0.49\textwidth}
    %\centering
    \includegraphics[width=0.99\linewidth,clip=true,trim=0.8cm 0.0cm 0.0cm 0.0cm]{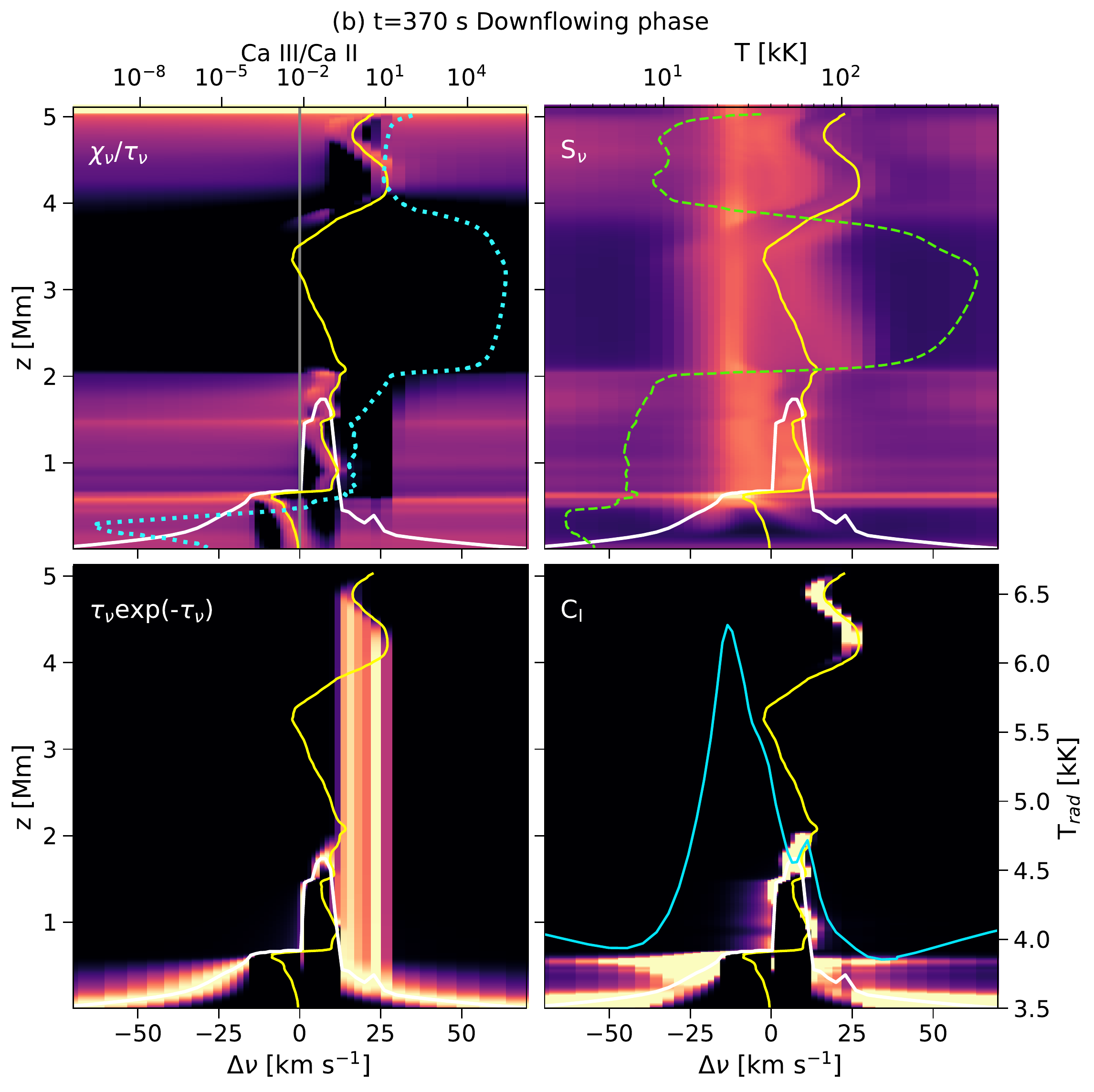} 
   % \caption{Formation of \cak{} for the downflowing phase.} 
    \label{figure:cont_Ca_DRRE} 
   % \vspace{4ex}
  \end{subfigure} 
  \begin{subfigure}[b]{0.49\textwidth}
    %\centering
    \includegraphics[width=0.99\linewidth,clip=true,trim=0.0cm 0.0cm 0.8cm 0.0cm]{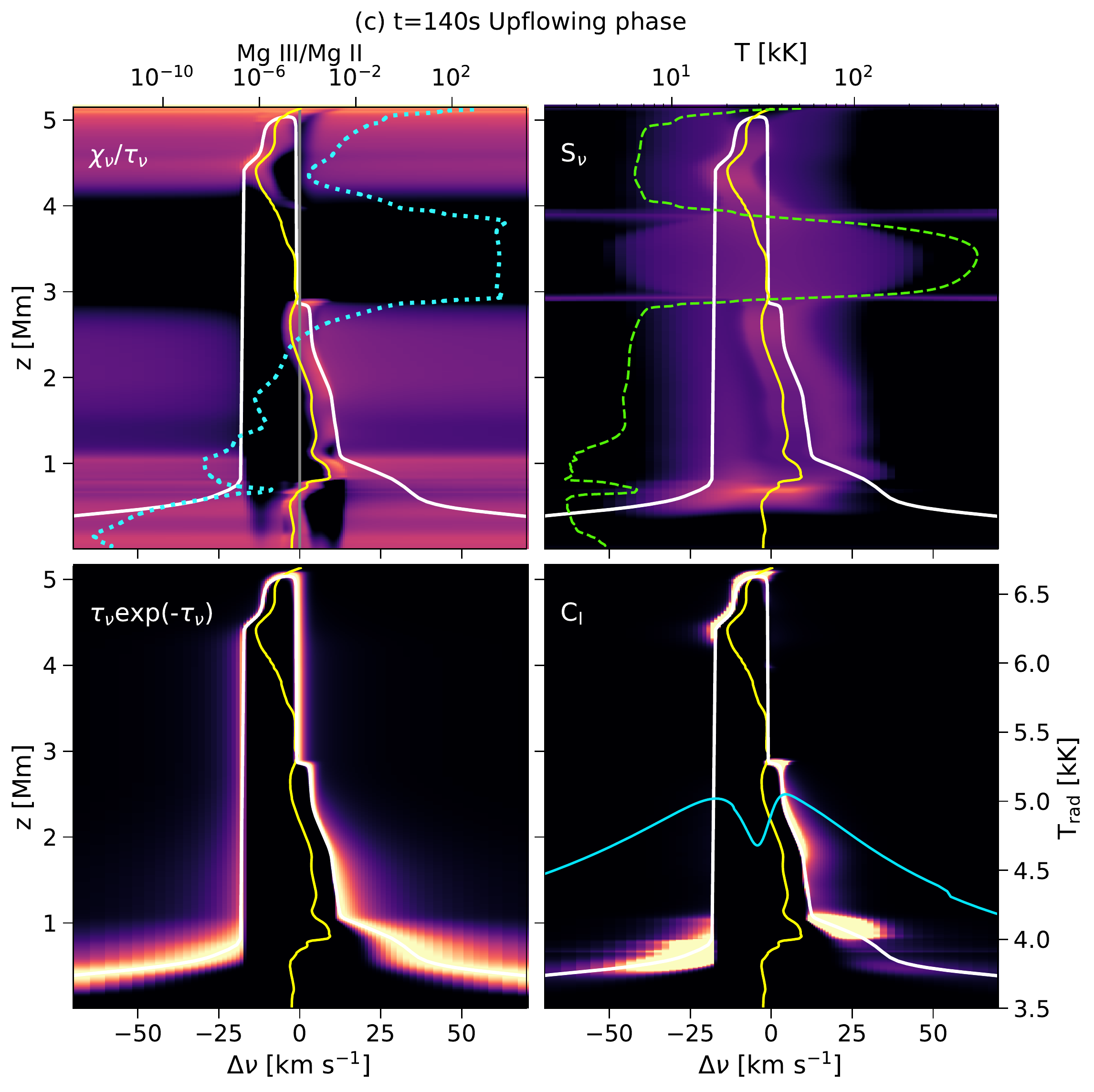} 
%    \caption{Formation of \Mgk{} for the upflowing phase.} 
    \label{figure:cont_Mg_RBE} 
  \end{subfigure}%%
  \begin{subfigure}[b]{0.49\textwidth}
    %\centering
    \includegraphics[width=0.99\linewidth,clip=true,trim=0.8cm 0.0cm 0.0cm 0.0cm]{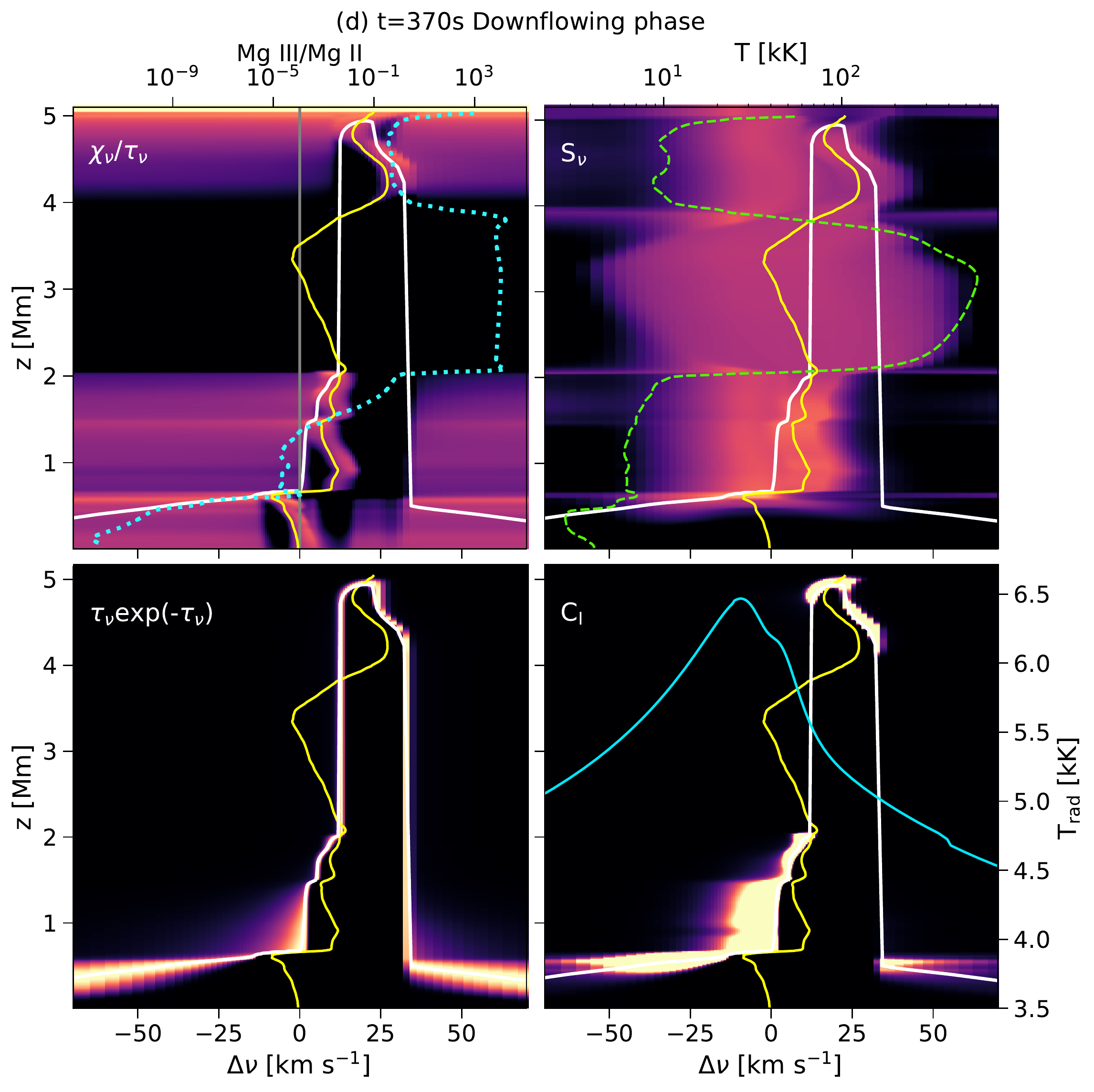} 
  %  \caption{Formation of \Mgk{} for the downflowing phase.} 
    \label{figure:cont_Mg_DRRE} 
  \end{subfigure} 
  \caption{Intensity formation breakdown figure for \cak{} and \Mgk{} spectral lines corresponding to spicule~1 from the model atmospheres at the location indicated by the vertical marker in Fig.~\ref{figure:synthetic_RBE_DRRE_Mg_Ca}. Subfigure~a: formation of \cak{} spectral line at t=140~s (upflowing phase) is depicted. Subfigure b: formation of \cak{} spectral line at t=370~s (downflowing phase) with a strong asymmetry towards the redward side of the line center is shown. Subfigure c: shows the formation of \Mgk{} line for the upflowing phase, and finally subfigure d: depicts the formation of a single peaked \Mgk{} spectral line for the downflowing phase. Each panel in the subfigures show (dark corresponds to low values and orange corresponds to high values) the quantity indicated at the top left corner as a function of geometric height $z$ and frequency from the line center in \kms{}. $\tau_{\nu}$=1 curve (white) and vertical velocity (yellow, positive is downflow) are overplotted in each panel in all the subfigures. The solid gray line (indicating 0 Doppler shift) is shown in the first panel for reference. In addition, the first panel shows the ionization fraction (dotted cyan line) of \ion{Ca}{II} (subfigures~a and b) and \ion{Mg}{II} (subfigures~c and d) as a function of $z$. The upper right panel also contains the plasma temperature (dashed green) as a function of $z$ in logarithmic scale specified along the top. The lower right panel also shows the specific intensity profiles (cyan), as radiation temperature with units specified on the right-hand side.}
  \label{figure:contribution_function_breakdown} 
\end{figure*}

%%--------Write below this line----------------
In Sect.~\ref{Subsubsection:Rad_transfer_diagnostics}, we saw that the Doppler shift of K$_{3}$ core in the synthetic \cak{} spectral line enhances or suppresses the respective K$_{2}$ peaks, but the effect may not be similar in \Mgk{}. For example, the synthetic \cak{} spectra for the upflowing phase shows a significant asymmetry in the K$_{2}$ peaks in Fig.~\ref{figure:synthetic_RBE_DRRE_Mg_Ca}, whereas the effect (though evident) is far less pronounced in the corresponding \Mgk{} spectra. Moreover, the downflowing phase shows a complete suppression of the \ion{Mg}{ii} k$_{2r}$ peak but we find a residual K$_{2r}$ in \ion{Ca}{ii} spectra. A similar remark can also be made from the observed downflowing RREs described in Sect.~\ref{Subsection:Spectral_signature}. These differences can be attributed to the fact that different spectral lines are sensitive to different regions in the solar atmosphere. We aim to understand the \cak{} and \Mgk{} spectral line formation observed in spicules based on contribution function analysis.

We used the method described in \citet{Carlsson_1997} to compute the contribution functions for \cak{} and \Mgk{} spectral lines formed during the upflowing and downflowing phases of spicules at the time steps and along the dashed vertical marker in Fig.~\ref{figure:synthetic_RBE_DRRE_Mg_Ca}. The contribution function $C_I$ was decomposed to the emergent intensity by

%\[ C_I(\nu) = \frac{dI(\nu)}{dz} = \frac{\chi(\nu)}{\tau(\nu)}\times S(\nu)\times \tau(\nu)\exp(-\tau(\nu)) \]
\[ C_I(\nu) = \frac{dI(\nu)}{dz} = \frac{\chi(\nu)}{\tau(\nu)}\times S(\nu)\times \tau(\nu)\, \mathrm{e}^{-\tau(\nu)} \]

into three parts $\chi$/$\tau$, $S$, and $\tau$ $\exp(-\tau)$ which are shown in subfigures~a, b, c, and d of Fig.~\ref{figure:contribution_function_breakdown} in the form a classical 4-panel diagram. The dependence on frequency (or wavelength) is shown along the $X$-axis, and the $Y$-axis shows the variation of the components along the geometric height scale $z$. On each panel we plot the vertical velocity as a function of geometric height (yellow solid) and optical depth unity as a function of frequency given by $\tau_{\nu}$=1 (white solid).

The first component shows the ratio between the total opacity $\chi$ and optical depth $\tau$. This is important when there are multiple emitting species (large $\chi$) at small optical depths. Moreover, this factor is also responsible for the asymmetry of spectral lines that is typical for atmospheres where there is a large velocity gradient (as is the case for spicules). In addition, we also show the ionization fraction of \ion{Ca}{II} to \ion{Ca}{III} as a function of height (dotted cyan).

The total (line plus continuum) source function ($S$) forms the second component that is strongly dependent on frequency for lines that are affected by PRD \citep[such as \cak{} and \Mgk{}, see][]{2013ApJ...772...90L,2018A&A...611A..62B}. Generally speaking, a large source function is indicative of strong emission in the lines. The temperature stratification is also shown in this panel (dashed green).

The third component, i.e. the factor $\tau$ $\exp(-\tau)$, generally highlights the region where $\tau_{\nu}$=1. Basically, it is high where the optical depth $\tau_{\nu}$=1.

Finally, the contribution function ($C_I$) is shown in the fourth panel along with the specific intensity of the spectral line (solid cyan) in radiation temperature units. 

Subfigures~a and b show the formation of \cak{} line during the upflowing and downflowing phases of spicule~1 corresponding to the location and the times indicated in Fig.~\ref{figure:synthetic_RBE_DRRE_Mg_Ca}. The upflowing phase shows two emission peaks in \cak{} that are asymmetric and the asymmetry is correlated with the shift of the K$_{3}$ core towards the blueward side of the line center. A close look at $C_I$ corresponding to the frequency of K$_{3}$ shows that the core is formed roughly around 5~Mm above the surface ($z$=0~Mm, defined by the height where $\tau$=1 for $\lambda=5000~\AA$) where the vertical velocity is approximately $-$10~\kms. This is clearly reflected in the Doppler shift of K$_{3}$ which is in tandem with the vertical velocity. The asymmetry in the K$_{2}$ peaks is caused by the Doppler shift of K$_{3}$ and a combination of velocities close to 0~\kms{} at 2.5~Mm and an upflow at 4.5~Mm that leads to a decrease in the radiation temperature of K$_{2v}$ (observed at $T_\mathrm{rad}\approx4.4$~kK) and an enhancement in K$_{2r}$ (that has a T$_\mathrm{rad}$ = 4.9~kK). The corresponding gas temperature at the height of formation of K$_{2r}$ is around 5.6~kK which is just 700~K higher than the radiation temperature. The panel corresponding to $\chi$/$\tau$ shows a large gap between 2.8--4.2~Mm in subfigure~a. This is mainly because of the lack of emitters (in this case \ion{Ca}{II} ions) at those heights, since all the \ion{Ca}{II} ions are converted to \ion{Ca}{III}. The temperature stratification plotted in the second panel supports this scenario since there is a sudden jump from 9~kK to 150~kK which is responsible for the ionization of singly ionized calcium atoms. We refer to this region as the line formation gap since there is no contribution to the emergent intensities from those heights. From the observable point-of-view, this gap is mainly due to the nonzero inclination of the spicule with respect to the LOS of the observer, and the limitation that the radiative transfer is performed along the vertical direction (dashed blue line in Fig.~\ref{figure:synthetic_RBE_DRRE_Mg_Ca}). In other words, regions surrounding the spicules are much hotter and have gas temperatures that can be as high as 150--1000~kK as shown in Fig.~\ref{figure:Context_v1_sim}. Consequently, the nonzero inclination causes the relatively cooler spicules to be sandwiched between the hot plasma region (around the spicule) to which the optically thick lines under investigation are not at all sensitive to. 

The downflowing phase shown in subfigure~b reveals an opposite and more complicated scenario. We find that the vertical velocity is predominantly positive (downflowing) except for heights below 0.6~Mm. Unlike the upflowing scenario, the $\tau_{\nu}$=1 has lower geometric height peaking merely at $z$=1.75~Mm, but there is a strong contribution between 4 and 5~Mm. The vertical velocity in this region is between 20--25~\kms, as also seen in the 2D velocity map shown in the bottom row of Fig.~\ref{figure:synthetic_RBE_DRRE_Mg_Ca}. This suggests that the top-most region, that would normally form the K$_{3}$ core (as is the case in panel a) is so strongly red shifted that it not only suppresses the K$_{2r}$, but also (nearly) envelops the far-wing K$_{1r}$. We also note that there is contribution around $\Delta \lambda$=25~\kms{} from regions below $z$=0.6~Mm. This contribution would likely lead to a normally occurring K$_{1r}$ with a depth closer to that of the blue wing K$_{1r}$. The tiny absorption dip at $\Delta \lambda$=10~\kms{} is due to contribution from the layers around $z$=1.8~Mm and forms a residual K$_{3}$ like feature. This aspect of an apparent K$_{1}$ being in fact strongly affected by the top-most layers of formation, normally responsible for a K$_{3}$ feature, is probably already the case for the representative profiles~3 and 8 in \citet{my_paper_3}, where the interplay of these classical features was studied, with K$_{3}$ being found to better reflect the top-most real velocities of the spicules. The formation gap due to complete ionization of \ion{Ca}{II} is bigger in this case since the inclination (relative to the same vertical line) is slightly smaller compared to the upflow.
%However, given the fact that the downflowing RRE in Fig.~\ref{figure:synthetic_RBE_DRRE_Mg_Ca} extends vertically up to \textasciitilde5~Mm, it is very likely that the contribution below 0.6~Mm is attributed to far-wings of \cak{} spectrum (likely K$_{1r}$), rather than K$_{3}$.
%The tiny emission at $\Delta \lambda$=15~\kms{} is due to contribution from the deeper layers around $z$=1.8~Mm. 

Subfigures~c and d show the formation of the \Mgk{} line for the same time instances as \cak{}. Though the formation mechanism is very similar to the latter, we do notice some differences between them. 
Firstly, in subfigure~c the asymmetry between the k$_{2}$ emission peaks is relatively low compared to subfigure~a. This is because though the velocity gradient between the formation of the two peaks is similar to the K$_{2}$ peaks, the Doppler shift of the k$_{3}$ core is smaller compared to K$_{3}$. This can be attributed to the difference in the vertical velocities around the region of formation between the two line cores with k$_{3}$ sampling slightly higher geometrical height than K$_{3}$. 
Secondly, the $\tau_{\nu}$=1 samples higher geometrical heights (with a peak of 5~Mm) during the downflowing phase in subfigure~d, unlike its \cak{} counterpart. This could be due to the higher opacity of magnesium in the upper layers of the solar atmosphere. The contribution function, however, bears resemblance to \cak{} with significant contribution between 4--5~Mm.
%Secondly, during the downflow shown in subfigure~d, the \ion{Mg}{ii} k$_{2r}$ peak disappears completely causing a significant asymmetry in the emergent spectral profile, that is not as prominent in \cak{}. This is because both $\tau_{\nu}$=1 formation height and contribution function are distinctly different from the downflowing phase in \cak{}. The k$_{3}$ core forms around a height of 5~Mm where the vertical velocity is significantly downflowing with an amplitude of 20--25~\kms{}. 
Finally, the \Mgk{} profiles are broader in comparison to the respective \cak{} profiles which is mainly because the Doppler width of spectral lines is inversely proportional to the square root of the mass (atomic weight) of the species under consideration; since calcium is roughly 1.7 times more massive than magnesium, the \Mgk{} profiles are broader. 
%In addition to the commonly known Doppler broadening, there could be spectral broadening due to unresolved non-thermal motions that affect the width of the spectral lines \citep[see, e.g.,][]{2003rtsa.book.....R}. Such treatments are beyond the scope of the current work. 

Apart from the above differences, the line formation mechanism and the contribution function analysis are similar in both spectral lines. Like subfigures~a and b, we see the presence of the formation gap in the $\chi$/$\tau$ and $C_I$ panels of subfigures~c and d, which is caused by the complete ionization of \ion{Mg}{II} to \ion{Mg}{III} due to high gas temperatures surrounding the spicule. During the downflowing phase we also see a complete suppression of the respective K$_{2r}$ (k$_{2r}$) peaks in the two cases. Moreover, the $T_\mathrm{rad}$ corresponding to the profiles in the downflowing phase is higher than the upflowing phase for both \cak{} and \Mgk{}, due to dissipation of currents from heating (like ambipolar diffusion) as described in Sect.~\ref{Subsubsection:heating_simulations}.

\section{Discussions}
\label{Section:Discussions}

\subsection{Multithermal nature of downflowing RREs}
\label{Subsection:multi_thermal_discussion}
This study aims to establish a link between the downflows found ubiquitously in the upper atmospheres of the Sun \citep[mainly the transition region and to some extent the lower corona, see][and the references therein]{1976ApJ...205L.177D,1981ApJ...251L.115G,1999ApJ...522.1148P,2012ApJ...749...60M} and the spicular downflows observed in the solar chromosphere \citep{2021A&A...647A.147B}. We followed an approach, where the spicular downflows were first detected in the high-resolution ground based datasets and the response of these downflows to the solar transition region and lower corona were studied. The reason for this approach is twofold: 1. Despite abundant observations of the transition region (and even lower coronal) downflows, their signatures have often remained elusive in the deeper layers of the Sun's atmosphere. Some studies, such as \citet{2012ApJ...749...60M}, found evidence of these downflows but they were mainly limited to hotter AIA channels (such as, 131, 171 and 193~\AA, with a minimum emission weighted temperature of 0.4~MK). Their signatures in the cooler channels, such as AIA 304~\AA, were found to be lacking. This could partially be attributable to the enormous difference in the plasma density between the (predominantly) chromospheric 304~\AA\ channel and the hotter (coronal) channels above it, which may cause a breakdown of these downflows as soon as they reach the chromosphere. 2. The superior spatial resolution of the ground-based telescopes enables the detection of small-scale features, such as spicules, which often remains undetected (or sometimes faintly detected) in the data from the space-based instruments. 
%This bottom-up approach, using coordinated observations, proved to be crucial for connecting these downflows across multithermal layers of the solar atmosphere.%Therefore, coordinated datasets from ground and space facilitates this bottom-up approach which proved to be crucial for connecting these downflows across multithermal layers of the solar atmosphere. 

We show several examples in this paper where the multithermal nature of the downflowing RREs is clearly revealed. They are found to have a broad temperature distribution, ranging from cooler chromospheric (\textasciitilde0.01~MK) to hotter transition and coronal temperatures. The term multithermal here refers to the fact that as the downflowing RREs evolve, different parts within the spicular structure are heated/cooled at different instances which causes them to appear across multiple wavelength channels (quasi)simultaneously sampling different regions of the solar atmosphere \citep[as observed in the case of outward propagating type-II spicules in][]{2011Sci...331...55D,Tiago_2014_heat,2019Sci...366..890S}. Occasionally, we looked at the signatures of downflows in the 131~\AA\ channel, if there was no prominent signal in the \Si{}~1400~\AA\ SJI (e.g. Fig.~\ref{figure:DRREs_X-t}). This lack of signal could either be due to the low exposure time (2~s) of the SJIs or the features could simply be too hot to be observed in the latter. However, given the visibility in the cooler 304~\AA\ channel, it is more likely that the signal was too faint in the SJIs. In some cases, such as the example shown in Fig.~\ref{figure:RBE_DRRE_Mg_Ca2}, we have also found signatures in the AIA 211~\AA\ channel (not shown) that has an emission weighted peak temperature of 2~MK, indicating that some of these downflows can be even hotter (though it is also possible that the emission can be due to the ions that are dominant at lower temperatures below 0.5~MK). Furthermore, the location of the hotter plasma with respect to the chromospheric downflows can often be slightly offset, indicating a scenario where the hottest emission lies on top of the cooler plasma. A precise cross-alignment among the SST, IRIS and SDO datasets ensures that the offsets seen in the different wavelength channels are likely due to different parts of the downflowing RREs being heated to different temperatures. Sometimes, a co-spatial strong emission in the hotter wavelength channels causes a lack of signature in the \halpha{} red wing images. This probably happens due to the lack of opacity in the chromosphere because of ionization of neutral hydrogen.

Though not the main focus of this paper, it is natural to ask what causes the spicules to have a multithermal nature and the driving mechanisms associated with them. We explain it briefly as follows. The heating to transition region and coronal temperatures can be explained from the perspective of the numerical simulation, where the spicules are launched due to a violent release in magnetic tension built up due to interaction between weak, granular fields and strong flux concentrations \citep{Juan_2017_Science}. Such a release of magnetic tension leads to a strong upward acceleration of the magnetized plasma, generation of transverse waves that propagate at Alfv\'enic speeds, along with electrical currents. Further, these currents are partially dissipated by the mechanism of ambipolar diffusion which heats the spicular plasma to at least transition region temperatures, whereas the remaining currents reach coronal heights where they are dissipated by Joule heating that heats up the loops associated with spicules to coronal temperatures \citep[to \textasciitilde2~MK. See][]{Juan_2017_Science,2017ApJ...845L..18D,Juan_2018} before falling back into the lower atmosphere. Observational evidence of this dissipation associated with spicules can be attributed to Kelvin-Helmholtz instabilities \citep{2018ApJ...856...44A} and/or resonant absorption associated with the expansion of strong-field magnetic field structures \citep{2019A&A...631A.105H}. In this paper, the focus is on the aftermaths or the returning phases of these spicules as they fall back into the chromosphere via the transition region and their impact in heating the chromosphere (discussed in Sect.~\ref{Subsection:simulation_discussion}). 
% This is suggestive of ionization of neutral hydrogen. 

The chromospheric and transition region spectra corresponding to the downflowing RREs show Doppler shifts ranging between 15--40~\kms{} redward of the respective line centers. This is consistent with the velocities that are commonly seen in the traditional RREs in the chromosphere \citep{Bart_3_motions,2013ApJ...769...44S,2015ApJ...802...26K,2021A&A...647A.147B}. We also observe that the velocity displacements of the K$_{3}$ (k$_{3}$) line cores of \cak{} (\Mgk{}) are in sync with the \halpha{} excursion asymmetries. This usually leads to a strong reduction (or sometimes even a complete disappearance) of one of the K$_{2}$ (k$_{2}$) peaks with an enhancement in the other. This has been reported in earlier studies such as \citet{Luc_2015} and \citet{my_paper_3}, with the latter finding evidences after characterizing millions of spectral profiles belonging to RBEs and RREs/downflowing RREs. The \Si{}~1393~\AA\ spectral profiles are noisier at these exposure times but a comparison of the $\lambda t$ diagrams reveal strong emission peaks that are in tandem with the chromospheric excursion asymmetries. Unfortunately, since the coverage of the IRIS rasters are limited in the spatial dimension, we do not observe the complete progression of the downflowing RREs in the IRIS spectroheliograms. We also find that none of the $\lambda t$ diagrams show preceding blue shifts that is characteristic to type-I spicules that are observed in the form of active region dynamic fibrils or quiet Sun mottles \citep{2008ApJ...673.1194L,2016ApJ...817..124S}. Moreover, the type-Is do not generally undergo rapid heating to transition region or coronal temperatures. This reinforces the proposition by \citet{2021A&A...647A.147B}, that the downflowing RREs are not likely associated with the return phases of type-Is. However, it is possible that the preceding upflow in this case is spatially displaced and is a part of large loop-like structure. Such a possibility is indicated in the numerical simulations by \citet{2020ApJ...889...95M} and will be discussed further in Sect.~\ref{subsection:discussion_loop_drre}. The apparent motion of the downflowing RREs range between 25--40~\kms{} that is compatible with the LOS velocities observed with the spectral lines formed in the upper atmosphere. This indicates that the downflowing brightenings seen in the transition region and coronal passbands are associated with real mass flows.  

The scenario presented in this paper aligns well with the scheme first proposed by \citet{1977A&A....55..305P} \citep[and later followed up by][]{1982ApJ...255..743A,1984ApJ...287..412A,2012ApJ...749...60M}, that the red shifts observed in the transition region spectral lines are caused by the emission from the returning phases of spicules that were formerly heated and/or injected into the solar corona. Using coronal spectroscopy, \citet{2012ApJ...749...60M} found that the dominant flow in the upper solar atmosphere changes its direction (from upflow to a predominating downflow) around a temperature of 1~MK. This would mean that the heated upflowing spicules undergo extensive radiative cooling and they later drain slowly back to the chromosphere via the transition region. These downflows have stronger emission and lower velocities in comparison to the upflows which could result in the net red shift observed in the transition region lines. Using coordinated ground and space-based datasets focusing on two different targets (one being an enhanced network and the other quiet Sun), we show unambiguously that the downflows are not just ubiquitous in the transition region, but they are equally prevalent in the solar chromosphere in the form of spicules. 

Despite providing a compelling case relating the transition region red shifts with spicules, it is challenging to estimate the spatio-temporal filling factor of these events. In other words, further studies are needed to quantify what fraction of these red shifts observed in the transition region are actually caused by the counterparts of downflowing RREs. Moreover, it is also known that the strength of the downflows in the transition region show regional differences. Relatively stronger flows have been observed in regions close to magnetic networks in comparison to inter-network regions (similar for a coronal hole), indicating that the strength of the magnetic field, along with the magnetic topology, might impact the measured red shifts \citep{2007ApJ...654..650M}. It is not fully clear how spicules would show such regional differences. It is possible that coronal loop properties might also play a role in modifying the transition red shifts in tandem with spicules but this is a speculation that needs further studies to reach a definitive consensus. Moreover, the coronal contribution associated with these downflowing features needs further investigation in the form of a statistical study \citep[like][]{Vasco_2016} to better understand (and firmly relate) the influence of downflowing RREs in the solar corona. Nevertheless, the results presented in this paper provide a crucial link for connecting the transition region and lower coronal downflows with their purely chromospheric counterparts, which was not observed in any of the earlier studies. The presence of such spicular downflows in the two datasets, with completely different magnetic field configurations, suggests that these downflows can be ubiquitously observed in the solar chromosphere. 

Though observations reveal that there are roughly equal number of red and blue excursion events from a chromospheric standpoint, it is possible that the emission spectra of the transition region lines are affected differently for different flows/excursions. \citet{2012ApJ...749...60M}, for example, found that rapidly upward propagating disturbances (with speeds between 50--150~\kms{}) contributed weakly to the emission profiles of the spectral lines formed at temperatures below 1~MK, whereas the radiatively cooled downflowing materials produced much more significant red wing emission that was thought to be responsible for the observed persistent red shifts in the transition region. Moreover, statistical analysis on over millions of \Mgk{} spectral profiles (based on $k$-means clustering) reveal that RBEs have weaker signatures in comparison to the red-shifted counterparts, both in terms of the Doppler shifts of the k$_{3}$ line cores and emission of k$_{2}$ peaks \citep[see Figs.~2 and 3 of][]{my_paper_3}. The rapidly upward propagating disturbances observed by \citet{2012ApJ...749...60M} were later linked to RBEs leading up to propagating coronal disturbances by combining numerical simulation and observations by \citet{2017ApJ...845L..18D}. Similarly, in this paper the idea is to suggest that the radiatively cooled downflowing plasma counterparts, speculated and observed by \citet{1977A&A....55..305P,1984ApJ...287..412A,2012ApJ...749...60M}, could be partially related to the downflowing RREs that were first reported by \citet{2021A&A...647A.147B}.

It is also interesting to note that not all rapid spicular downflows observed in the chromosphere show coronal or transition region responses. Since the downflowing RREs are most likely the returning counterparts of the previously heated spicular upflows, it is possible that many such upflows are simply not hot enough to be visible in the coronal or even transition region passbands in the first place. We also note that a significant fraction of the heating associated with spicules is not local, but distributed along the coronal loops \citep{Juan_2018}, that might prevent us from detecting such events in the hotter channels. In fact, \citet{Vasco_2016} statistically established a lower limit of only 6\% match between the brightenings associated in the lower corona with the rapid upflows (RBEs and RREs) seen in quiet Sun \halpha{} observations. This indicates that a large number of such events are either not hot enough to be detected in the passbands sampling at least 0.8~MK, or too small to be resolved with current space-based instruments. As a result, a fraction of the ubiquitous chromospheric spicular downflows observed in the Sun could be attributable to the return flows that trace the path of the preceding type-II spicules, which are hot enough to cause partial ionization of neutral hydrogen at about 0.01~MK, but not hot enough to appear in the lower coronal passbands \citep{2019A&A...632A..96R}.

\subsection{Impact on heating during the downflow}
\label{Subsection:simulation_discussion}
The spectra corresponding to the maximum excursion of the downflowing RREs presented in Sect.~\ref{Subsection:Spectral_signature} show that they have enhanced emission peaks in the \cak{} and \Mgk{} spectral lines and are broader when compared to the average profiles. This is in support of the results presented earlier in \citet{my_paper_3}, where the RRE/downflowing RRE-like profiles are enhanced and broader in comparison to the RBEs. This hinted at the possibility of heating-based mechanisms during the downflows. Forward modeling on a state-of-the-art MHD simulation of spicules, shows that the \cak{} and \Mgk{} synthetic profiles are similar to what is found in the observations. This is shown in Fig.~\ref{figure:synthetic_RBE_DRRE_Mg_Ca}.

A detailed analysis of the temperature, internal energy, and various heating mechanisms corresponding to the type-II spicules revealed that an enhancement in heating is observed during the downflowing phase which is well correlated with the marked rise in the gas temperature (by roughly 2--3~kK) of the spicules. In fact, both temperature and heating (normalized by the number of particles) starts showing enhancement right around the time when the downflow initiates (see Fig.~\ref{figure:heating_per_particle}). We did not include heating due to adiabatic compression in our analysis because we wanted to eliminate the possibility of a rise in temperature without any exchange of heat between the spicule and the surroundings. Most of the enhancement in heating towards the downflowing phase comes from ambipolar diffusion which was a crucial ingredient in the simulation by \cite{Juan_2017_Science, Juan_2018}. The increase in ambipolar diffusion is due to the decrease in the plasma density which causes the collision frequency between ions and neutrals to drop. This further results in dissipation or damping of Alfv\'enic waves and generation of currents that heats the plasma \citep{Juan_2017_Science}. The internal energy, on the other hand, shows no significant changes during the evolution of the spicules because as the temperature rises, the heated plasma becomes less dense. A detailed study of the line formation mechanism using contribution functions indicate that the chromospheric \cak{} and \Mgk{} lines are sensitive to the region of the solar atmosphere where this heating takes place, which, in turn, is reflected in the emergent intensities.
%Furthermore, a line formation gap is also found to be associated with spicules mainly because the cooler spicular material is surrounded by hotter plasma where the singly ionized calcium and magnesium atoms are completely ionized. 
%1. lifetime is comparable to observations ~600s. 

\subsection{Downflowing RRE along a loop}
\label{subsection:discussion_loop_drre}
\begin{figure*}[h]
   \centering
   \includegraphics[width=\textwidth]{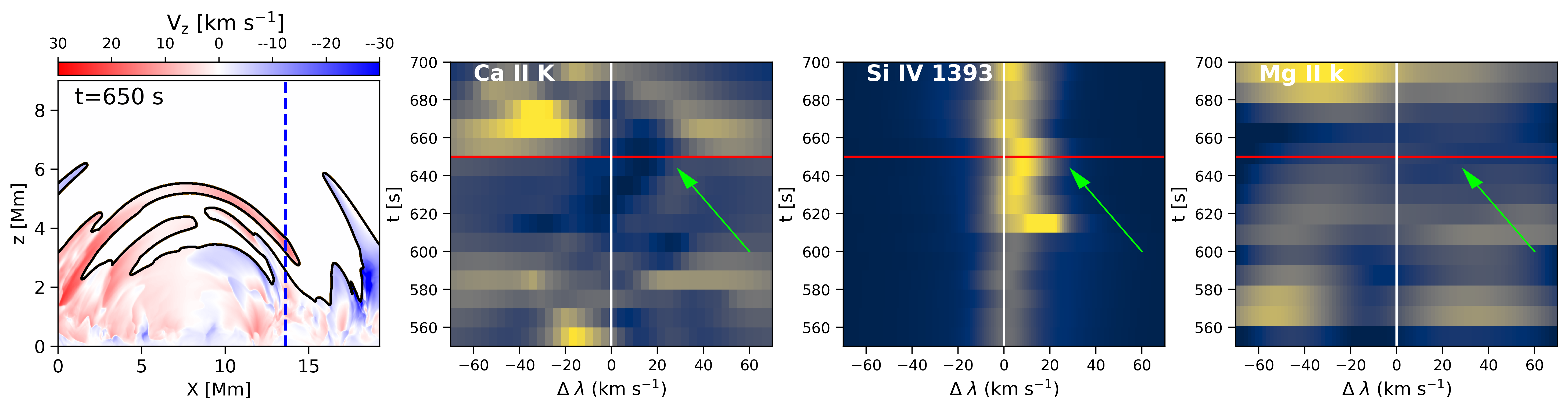}
   \caption{Scenario outlining an alternative mechanism of observing downflowing RREs along a loop-like structure from a simulation perspective. Panels from left to right show: a cut-out of the 2D vertical velocity (v$_{\mathrm{z}}$) map focusing on the downflowing RRE formed along the loop, $\lambda t$ diagrams in \cak{}, \Si{}~1393~\AA, and \Mgk{}, respectively, corresponding to the location marked with a blue dashed vertical line in the v$_{\mathrm{z}}$ map (leftmost panel). The v$_{\mathrm{z}}$ map corresponds to the region-of-interest shown in the bottom row of Fig.~\ref{figure:Context_v1_sim} at $t$=650~s. Each of the $\lambda t$ diagrams show the location of Doppler shift $\Delta \lambda=0$ (indicated with a white solid line), the time step (in red horizontal line), and an arrow pointing at the time interval corresponding to the maximum redward excursions of the respective line centers during the evolution of the downflowing RRE. An animation of this figure is available at \url{https://www.dropbox.com/s/h7ib86waeweh25t/Loop_DRRE.mp4?dl=0}.
   } 
        \label{figure:Loop_DRRE}%
    \end{figure*}

Towards the end of Sect.~\ref{Subsection:TR_coronal_response}, and later in Sect.~\ref{Subsection:multi_thermal_discussion}, it was indicated that the downflowing phases of RREs could be spatially displaced from their upflowing phases and be formed along loop-like structures \citep[see][who proposed it on the basis of their numerical simulations]{2020ApJ...889...95M}. Such a mechanism would be an alternative to the examples presented in Fig.~\ref{figure:heating_per_particle}, where the downflow and the upflow are nearly co-located. While analyzing the chromospheric counterparts of the low-lying transition region loops, \citet{2018A&A...611L...6P} found that spicules are predominant and often lie close to (or share) the footpoints of such loops, and have strong Doppler shifts. However, most of the plasma associated with such loops were found to be heated to at least transition region temperatures, with no distinct signature in the chromosphere. In this section, we investigate and discuss the possibility of observing downflowing RREs forming along loops in the context of the numerical simulation described in this paper.

The 2D slice of the vertical velocity shown in Fig.~\ref{figure:Loop_DRRE}, and its associated animation, clearly outlines the scenario where we find a downflowing RRE associated with a loop that originates as a spicular upflow roughly at $x$= 3~Mm, and nearly 400~s before the indicated snapshot at $t$=650~s. For the sake of brevity, we limit the duration of the event focusing only on the downflowing phase of spicular loop in Fig.~\ref{figure:Loop_DRRE}. However, we encourage the readers to view the animation associated with Fig.~\ref{figure:Context_v1_sim} from time $t$ = 240~s to 700~s to fully appreciate its progression from the blueward to redward excursion phase. 

The $\lambda t$ diagrams for the synthetic \cak{}, \Si{}~1393~\AA\ and \Mgk{} spectral lines, corresponding to the location of the dashed vertical marker, show asymmetries towards the redward side of the respective line centers that are in tandem with the observations presented in Figs.~\ref{figure:RBE_DRRE_Mg_Ca1}, \ref{figure:RBE_DRRE_Mg_Ca2}, and \ref{figure:RBE_DRRE_Mg_Ca3}. Moreover, the synthetic $\lambda t$ diagrams lack preceding blueward excursions (similar to the observed $\lambda t$ diagrams) that excludes them from being related to type-I spicules. A quasi-temporal redward excursion in all the three spectral lines strongly indicates that the loop-like structures observed in the chromosphere and the transition region could also lead to downflowing RREs -- that are multithermal by nature and can be heated to 0.08~MK, similar to the spectral signatures found in the coordinated observations. Moreover, the heating associated with these loop-like downflowing RREs are found to be governed by a mechanism that is similar to examples presented in Sect.~\ref{Subsubsection:heating_simulations}. 

The scenario presented in this section is slightly different from the observations by \citet{2018A&A...611L...6P}, who found subtle or limited traces of the low-lying loops in the chromosphere. Moreover, they indicated that the footpoints of those loops were close to the footpoints of the spicules, but they were clearly distinct from one another. However, here we discussed a possibility where spicules (downflowing RREs or RBEs) are basically formed along the structure of the loop itself, and they show clear evidences of strong Doppler shifts both in the chromosphere and the transition region. 

\section{Concluding remarks}
\label{Section:Conclusion}

We study the transition region and coronal response of the recently reported downflowing RREs. Coordinated ground and space-based observations reveal that these downflows are ubiquitous, have a multithermal nature with a broad temperature distribution, and can be heated to at least transition region temperatures. The downflows are most likely a result of previously heated spicular material that underwent cooling and eventually fell back into the chromosphere -- a scenario that was already outlined by studies such as \citet{1977A&A....55..305P,1984ApJ...287..412A} and \citet{2012ApJ...749...60M}. The current work provides the missing link between the abundant downflows observed in the transition region (and lower corona) with their chromospheric counterparts that has so far remained elusive. Although it is hard to estimate what fraction of the transition region red shifts are directly associated with downflowing RREs for the entire solar atmosphere, the analysis presented in this study provides a compelling case that links these observed red shifts with spicules in the chromosphere. Moreover, the Doppler shifts of the \Si{} spectral line are close to the observed average red shifts found in the transition region that leads us to believe that the downflows might play an important role in explaining the observed red shifts. 

%%The preceding discussion indicates that the line formation mechanism for features that have strong gradient in the LOS velocity can be rather non-trivial. 
A detailed comparison with the numerical simulation from \citet{2017ApJ...847...36M,Juan_2017_Science} successfully reproduced the observed spectral properties of the downflowing RREs to a large extent. Analysis based on contribution functions, and reinforcing the investigation of \citet{my_paper_3}, shows that care should be taken when interpreting the positions of classical spectral features such as K$_{1}$ and K$_{3}$. Moreover, an alternate scenario has been presented from the perspective of numerical simulations that raises the possibility of observing downflowing RREs along loop-like structures and could explain the lack of preceding blueward shifts in the spectral time diagrams. It could be interesting to look for such examples in future studies from coordinated high-resolution observations from ground and space-based facilities. We also showed that the returning phases of type-II spicules are heated in comparison to the upflowing phases. This heating was due to a combination of various mechanisms, but ambipolar diffusion was found to play a pivotal role that led to an enhancement in the temperature by about 2--3~kK. This is further observed in the form of enhanced emission in the synthetic \cak{} and \Mgk{} spectra, which seems to explain the enhancement for the observed \cak{} and \Mgk{} spectra in the examples shown in this paper and also in \citet{my_paper_3}. 

Finally, we note that the effect of non-equilibrium ionization of different atomic species, such as, hydrogen, helium or silicon, has not been included in the numerical simulation used in this paper. Non-equilibrium ionization can cause an increase in the number density of electrons in the upper chromosphere due to longer recombination timescales \citep{2020ApJ...889...95M}. Therefore, any heating (due to ambipolar diffusion, for example) will directly impact the temperature instead of ionizing/recombining the plasma, producing multithermal structures with even sharper temperature variations \citep{2021ApJ...906...82C}.  A detailed study including the effects of non-equilibrium ionization is currently underway and will form a part of a follow-up paper. 

% line formation mechanisms, especially for features that have strong gradients in the LOS velocity, can be rather non-trivial to interpret.

\begin{acknowledgements}
The Swedish 1-m Solar Telescope is operated on the island of La Palma
by the Institute for Solar Physics of Stockholm University in the
Spanish Observatorio del Roque de los Muchachos of the Instituto de
Astrof{\'\i}sica de Canarias.
The Institute for Solar Physics is supported by a grant for research infrastructures of national importance from the Swedish Research Council (registration number 2017-00625).
IRIS is a NASA small explorer mission developed and operated by LMSAL with mission operations executed at NASA Ames Research Center and major contributions to downlink communications funded by ESA and the Norwegian Space Centre.
% SDO?
%
This research is supported by the Research Council of Norway, project number 250810, and through its Centres of Excellence scheme, project number 262622, and through grants of computing time from the Programme for Supercomputing.
VMJH is supported by the European Research Council (ERC) under the European Union’s Horizon 2020 research and innovation programme (SolarALMA, grant agreement No. 682462).
DNS acknowledges support by the Synergy Grant number 810218 (ERC-2018-SyG) of the European Research Council,
and the project PGC2018-095832-B-I00 of the the Spanish Ministry of Science, Innovation and Universities.
JMS gratefully acknowledges support by NASA grants 80NSSC18K1285, 80NSSC21K0737 and NNG09FA40C (IRIS). The simulations have been run on clusters from the Notur project, and the Pleiades cluster through the computing project s1061, s1630, and s2053, from the High End Computing (HEC) division of NASA.
BDP is supported by NASA contract NNG09FA40C (IRIS).
We are indebted to Rob Rutten for his excellent routines for alignment of SDO and ground-based data. %, which he has kindly made public, and support thereof. 
% 25 May 2017 : Ainar?
We are thankful for the assistance of Ainar Drews with the observations of the 25~May~2017 set.
%
% 20 Sep 2020 : Pit service mode
We thank Pit S\"utterlin for excellent SST service mode observations during the 2020 COVID-19 season. 
We made much use of NASA's Astrophysics Data System Bibliographic Services.
Parts of the results in this work make use of the colormaps in the CMasher package \citep{2020JOSS....5.2004V}.
\end{acknowledgements}

% WARNING
%-------------------------------------------------------------------
% Please note that we have included the references to the file aa.dem in
% order to compile it, but we ask you to:
%
% - use BibTeX with the regular commands:
%   \bibliographystyle{aa} % style aa.bst
%   \bibliography{Yourfile} % your references Yourfile.bib
%
% - join the .bib files when you upload your source files
%-------------------------------------------------------------------
\bibliographystyle{aa} % style aa.bst
\bibliography{spicules_5.bib}

\begin{thebibliography}{80}
\expandafter\ifx\csname natexlab\endcsname\relax\def\natexlab#1{#1}\fi

\bibitem[{{Antolin} {et~al.}(2018){Antolin}, {Schmit}, {Pereira}, {De Pontieu},
  \& {De Moortel}}]{2018ApJ...856...44A}
{Antolin}, P., {Schmit}, D., {Pereira}, T.~M.~D., {De Pontieu}, B., \& {De
  Moortel}, I. 2018, \apj, 856, 44

\bibitem[{{Asplund} {et~al.}(2009){Asplund}, {Grevesse}, {Sauval}, \&
  {Scott}}]{2009ARA&A..47..481A}
{Asplund}, M., {Grevesse}, N., {Sauval}, A.~J., \& {Scott}, P. 2009, \araa, 47,
  481

\bibitem[{{Athay}(1984)}]{1984ApJ...287..412A}
{Athay}, R.~G. 1984, \apj, 287, 412

\bibitem[{{Athay} \& {Holzer}(1982)}]{1982ApJ...255..743A}
{Athay}, R.~G. \& {Holzer}, T.~E. 1982, \apj, 255, 743

\bibitem[{{Bj{\o}rgen} {et~al.}(2018){Bj{\o}rgen}, {Sukhorukov}, {Leenaarts},
  {Carlsson}, {de la Cruz Rodr{\'\i}guez}, {Scharmer}, \&
  {Hansteen}}]{2018A&A...611A..62B}
{Bj{\o}rgen}, J.~P., {Sukhorukov}, A.~V., {Leenaarts}, J., {et~al.} 2018, \aap,
  611, A62

\bibitem[{{Bose} {et~al.}(2019{\natexlab{a}}){Bose}, {Henriques}, {Joshi}, \&
  {Rouppe van der Voort}}]{my_paper_3}
{Bose}, S., {Henriques}, V. M.~J., {Joshi}, J., \& {Rouppe van der Voort}, L.
  2019{\natexlab{a}}, \aap, 631, L5

\bibitem[{{Bose} {et~al.}(2019{\natexlab{b}}){Bose}, {Henriques}, {Rouppe van
  der Voort}, \& {Pereira}}]{Bose_2019}
{Bose}, S., {Henriques}, V. M.~J., {Rouppe van der Voort}, L., \& {Pereira}, T.
  M.~D. 2019{\natexlab{b}}, \aap, 627, A46

\bibitem[{{Bose} {et~al.}(2021){Bose}, {Joshi}, {Henriques}, \& {van der
  Voort}}]{2021A&A...647A.147B}
{Bose}, S., {Joshi}, J., {Henriques}, V. M.~J., \& {van der Voort}, L.~R. 2021,
  \aap, 647, A147

\bibitem[{{Brekke} {et~al.}(1997){Brekke}, {Hassler}, \&
  {Wilhelm}}]{1997SoPh..175..349B}
{Brekke}, P., {Hassler}, D.~M., \& {Wilhelm}, K. 1997, \solphys, 175, 349

\bibitem[{{Carlsson} \& {Leenaarts}(2012)}]{2012A&A...539A..39C}
{Carlsson}, M. \& {Leenaarts}, J. 2012, \aap, 539, A39

\bibitem[{{Carlsson} \& {Stein}(1997)}]{Carlsson_1997}
{Carlsson}, M. \& {Stein}, R.~F. 1997, \apj, 481, 500

\bibitem[{{Chintzoglou} {et~al.}(2021){Chintzoglou}, {De Pontieu},
  {Mart{\'\i}nez-Sykora}, {Hansteen}, {de la Cruz Rodr{\'\i}guez},
  {Szydlarski}, {Jafarzadeh}, {Wedemeyer}, {Bastian}, \& {Sainz
  Dalda}}]{2021ApJ...906...82C}
{Chintzoglou}, G., {De Pontieu}, B., {Mart{\'\i}nez-Sykora}, J., {et~al.} 2021,
  \apj, 906, 82

\bibitem[{{Dadashi} {et~al.}(2011){Dadashi}, {Teriaca}, \&
  {Solanki}}]{2011A&A...534A..90D}
{Dadashi}, N., {Teriaca}, L., \& {Solanki}, S.~K. 2011, \aap, 534, A90

\bibitem[{{de la Cruz Rodr{\'\i}guez} {et~al.}(2015){de la Cruz
  Rodr{\'\i}guez}, {Hansteen}, {Bellot-Rubio}, \&
  {Ortiz}}]{2015ApJ...810..145D}
{de la Cruz Rodr{\'\i}guez}, J., {Hansteen}, V., {Bellot-Rubio}, L., \&
  {Ortiz}, A. 2015, \apj, 810, 145

\bibitem[{{De Pontieu} {et~al.}(2012){De Pontieu}, {Carlsson}, {Rouppe van der
  Voort}, {Rutten}, {Hansteen}, \& {Watanabe}}]{Bart_3_motions}
{De Pontieu}, B., {Carlsson}, M., {Rouppe van der Voort}, L.~H.~M., {et~al.}
  2012, \apjl, 752, L12

\bibitem[{{De Pontieu} {et~al.}(2017{\natexlab{a}}){De Pontieu}, {De Moortel},
  {Martinez-Sykora}, \& {McIntosh}}]{2017ApJ...845L..18D}
{De Pontieu}, B., {De Moortel}, I., {Martinez-Sykora}, J., \& {McIntosh}, S.~W.
  2017{\natexlab{a}}, \apjl, 845, L18

\bibitem[{{De Pontieu} {et~al.}(2017{\natexlab{b}}){De Pontieu},
  {Mart{\'{\i}}nez-Sykora}, \& {Chintzoglou}}]{2017ApJ...849L...7D}
{De Pontieu}, B., {Mart{\'{\i}}nez-Sykora}, J., \& {Chintzoglou}, G.
  2017{\natexlab{b}}, \apjl, 849, L7

\bibitem[{{De Pontieu} {et~al.}(2007){De Pontieu}, {McIntosh}, {Hansteen},
  {Carlsson}, {Schrijver}, {Tarbell}, {Title}, {Shine}, {Suematsu}, \&
  {Tsuneta}}]{Bart_2007_PASJ}
{De Pontieu}, B., {McIntosh}, S., {Hansteen}, V.~H., {et~al.} 2007, \pasj, 59,
  S655

\bibitem[{{De Pontieu} {et~al.}(2011){De Pontieu}, {McIntosh}, {Carlsson},
  {Hansteen}, {Tarbell}, {Boerner}, {Martinez-Sykora}, {Schrijver}, \&
  {Title}}]{2011Sci...331...55D}
{De Pontieu}, B., {McIntosh}, S.~W., {Carlsson}, M., {et~al.} 2011, Science,
  331, 55

\bibitem[{{De Pontieu} {et~al.}(2009){De Pontieu}, {McIntosh}, {Hansteen}, \&
  {Schrijver}}]{2009ApJ...701L...1D}
{De Pontieu}, B., {McIntosh}, S.~W., {Hansteen}, V.~H., \& {Schrijver}, C.~J.
  2009, \apjl, 701, L1

\bibitem[{{De Pontieu} {et~al.}(2014){De Pontieu}, {Title}, {Lemen}, {Kushner},
  {Akin}, {Allard}, {Berger}, {Boerner}, {Cheung}, {Chou}, {Drake}, {Duncan},
  {Freeland}, {Heyman}, {Hoffman}, {Hurlburt}, {Lindgren}, {Mathur}, {Rehse},
  {Sabolish}, {Seguin}, {Schrijver}, {Tarbell}, {W{\"u}lser}, {Wolfson},
  {Yanari}, {Mudge}, {Nguyen-Phuc}, {Timmons}, {van Bezooijen}, {Weingrod},
  {Brookner}, {Butcher}, {Dougherty}, {Eder}, {Knagenhjelm}, {Larsen},
  {Mansir}, {Phan}, {Boyle}, {Cheimets}, {DeLuca}, {Golub}, {Gates}, {Hertz},
  {McKillop}, {Park}, {Perry}, {Podgorski}, {Reeves}, {Saar}, {Testa}, {Tian},
  {Weber}, {Dunn}, {Eccles}, {Jaeggli}, {Kankelborg}, {Mashburn}, {Pust},
  {Springer}, {Carvalho}, {Kleint}, {Marmie}, {Mazmanian}, {Pereira}, {Sawyer},
  {Strong}, {Worden}, {Carlsson}, {Hansteen}, {Leenaarts}, {Wiesmann},
  {Aloise}, {Chu}, {Bush}, {Scherrer}, {Brekke}, {Martinez-Sykora}, {Lites},
  {McIntosh}, {Uitenbroek}, {Okamoto}, {Gummin}, {Auker}, {Jerram}, {Pool}, \&
  {Waltham}}]{Bart2014}
{De Pontieu}, B., {Title}, A.~M., {Lemen}, J.~R., {et~al.} 2014, \solphys, 289,
  2733

\bibitem[{{Dere}(1982)}]{1982SoPh...77...77D}
{Dere}, K.~P. 1982, \solphys, 77, 77

\bibitem[{{Dere} {et~al.}(2009){Dere}, {Landi}, {Young}, {Del Zanna},
  {Landini}, \& {Mason}}]{2009A&A...498..915D}
{Dere}, K.~P., {Landi}, E., {Young}, P.~R., {et~al.} 2009, \aap, 498, 915

\bibitem[{{Doschek} {et~al.}(1976){Doschek}, {Feldman}, \&
  {Bohlin}}]{1976ApJ...205L.177D}
{Doschek}, G.~A., {Feldman}, U., \& {Bohlin}, J.~D. 1976, \apjl, 205, L177

\bibitem[{{Gebbie} {et~al.}(1981){Gebbie}, {Hill}, {November}, {Gurman},
  {Shine}, {Woodgate}, {Athay}, {Tandberg-Hanssen}, {Toomre}, \&
  {Simon}}]{1981ApJ...251L.115G}
{Gebbie}, K.~B., {Hill}, F., {November}, L.~J., {et~al.} 1981, \apjl, 251, L115

\bibitem[{{Gudiksen} {et~al.}(2011){Gudiksen}, {Carlsson}, {Hansteen}, {Hayek},
  {Leenaarts}, \& {Mart{\'\i}nez-Sykora}}]{2011A&A...531A.154G}
{Gudiksen}, B.~V., {Carlsson}, M., {Hansteen}, V.~H., {et~al.} 2011, \aap, 531,
  A154

\bibitem[{{Hansteen}(1993)}]{1993ApJ...402..741H}
{Hansteen}, V. 1993, \apj, 402, 741

\bibitem[{{Hansteen} {et~al.}(2010){Hansteen}, {Hara}, {De Pontieu}, \&
  {Carlsson}}]{2010ApJ...718.1070H}
{Hansteen}, V.~H., {Hara}, H., {De Pontieu}, B., \& {Carlsson}, M. 2010, \apj,
  718, 1070

\bibitem[{{Henriques}(2012)}]{2012A&A...548A.114H}
{Henriques}, V.~M.~J. 2012, \aap, 548, A114

\bibitem[{{Henriques} {et~al.}(2016){Henriques}, {Kuridze}, {Mathioudakis}, \&
  {Keenan}}]{Vasco_2016}
{Henriques}, V.~M.~J., {Kuridze}, D., {Mathioudakis}, M., \& {Keenan}, F.~P.
  2016, \apj, 820, 124

\bibitem[{{Howson} {et~al.}(2019){Howson}, {De Moortel}, {Antolin}, {Van
  Doorsselaere}, \& {Wright}}]{2019A&A...631A.105H}
{Howson}, T.~A., {De Moortel}, I., {Antolin}, P., {Van Doorsselaere}, T., \&
  {Wright}, A.~N. 2019, \aap, 631, A105

\bibitem[{{Kjeldseth-Moe} {et~al.}(1988){Kjeldseth-Moe}, {Brynildsen},
  {Brekke}, {Engvold}, {Maltby}, {Bartoe}, {Brueckner}, {Cook}, {Dere}, \&
  {Socker}}]{1988ApJ...334.1066K}
{Kjeldseth-Moe}, O., {Brynildsen}, N., {Brekke}, P., {et~al.} 1988, \apj, 334,
  1066

\bibitem[{{Kuridze} {et~al.}(2015){Kuridze}, {Henriques}, {Mathioudakis},
  {Erd{\'e}lyi}, {Zaqarashvili}, {Shelyag}, {Keys}, \&
  {Keenan}}]{2015ApJ...802...26K}
{Kuridze}, D., {Henriques}, V., {Mathioudakis}, M., {et~al.} 2015, \apj, 802,
  26

\bibitem[{{Langangen} {et~al.}(2008{\natexlab{a}}){Langangen}, {Carlsson},
  {Rouppe van der Voort}, {Hansteen}, \& {De Pontieu}}]{2008ApJ...673.1194L}
{Langangen}, {\O}., {Carlsson}, M., {Rouppe van der Voort}, L., {Hansteen}, V.,
  \& {De Pontieu}, B. 2008{\natexlab{a}}, \apj, 673, 1194

\bibitem[{{Langangen} {et~al.}(2008{\natexlab{b}}){Langangen}, {De Pontieu},
  {Carlsson}, {Hansteen}, {Cauzzi}, \& {Reardon}}]{2008ApJ...679L.167L}
{Langangen}, {\O}., {De Pontieu}, B., {Carlsson}, M., {et~al.}
  2008{\natexlab{b}}, \apjl, 679, L167

\bibitem[{{Leenaarts} {et~al.}(2012){Leenaarts}, {Pereira}, \&
  {Uitenbroek}}]{2012A&A...543A.109L}
{Leenaarts}, J., {Pereira}, T., \& {Uitenbroek}, H. 2012, \aap, 543, A109

\bibitem[{{Leenaarts} {et~al.}(2013){Leenaarts}, {Pereira}, {Carlsson},
  {Uitenbroek}, \& {De Pontieu}}]{2013ApJ...772...90L}
{Leenaarts}, J., {Pereira}, T.~M.~D., {Carlsson}, M., {Uitenbroek}, H., \& {De
  Pontieu}, B. 2013, \apj, 772, 90

\bibitem[{{Lemen} {et~al.}(2012){Lemen}, {Title}, {Akin}, {Boerner}, {Chou},
  {Drake}, {Duncan}, {Edwards}, {Friedlaender}, {Heyman}, {Hurlburt}, {Katz},
  {Kushner}, {Levay}, {Lindgren}, {Mathur}, {McFeaters}, {Mitchell}, {Rehse},
  {Schrijver}, {Springer}, {Stern}, {Tarbell}, {Wuelser}, {Wolfson}, {Yanari},
  {Bookbinder}, {Cheimets}, {Caldwell}, {Deluca}, {Gates}, {Golub}, {Park},
  {Podgorski}, {Bush}, {Scherrer}, {Gummin}, {Smith}, {Auker}, {Jerram},
  {Pool}, {Soufli}, {Windt}, {Beardsley}, {Clapp}, {Lang}, \&
  {Waltham}}]{2012SoPh..275...17L}
{Lemen}, J.~R., {Title}, A.~M., {Akin}, D.~J., {et~al.} 2012, \solphys, 275, 17

\bibitem[{{L{\"o}fdahl} {et~al.}(2018){L{\"o}fdahl}, {Hillberg}, {de la Cruz
  Rodriguez}, {Vissers}, {Scharmer}, {Hagfors Haugan}, \&
  {Fredvik}}]{2018arXiv180403030L}
{L{\"o}fdahl}, M.~G., {Hillberg}, T., {de la Cruz Rodriguez}, J., {et~al.}
  2018, arXiv e-prints, arXiv:1804.03030

\bibitem[{{Mariska}(1987)}]{1987ApJ...319..465M}
{Mariska}, J.~T. 1987, \apj, 319, 465

\bibitem[{{Mart{\'\i}nez-Sykora}
  {et~al.}(2017{\natexlab{a}}){Mart{\'\i}nez-Sykora}, {De Pontieu}, {Carlsson},
  {Hansteen}, {N{\'o}brega-Siverio}, \& {Gudiksen}}]{2017ApJ...847...36M}
{Mart{\'\i}nez-Sykora}, J., {De Pontieu}, B., {Carlsson}, M., {et~al.}
  2017{\natexlab{a}}, \apj, 847, 36

\bibitem[{{Mart{\'\i}nez-Sykora} {et~al.}(2018){Mart{\'\i}nez-Sykora}, {De
  Pontieu}, {De Moortel}, {Hansteen}, \& {Carlsson}}]{Juan_2018}
{Mart{\'\i}nez-Sykora}, J., {De Pontieu}, B., {De Moortel}, I., {Hansteen},
  V.~H., \& {Carlsson}, M. 2018, \apj, 860, 116

\bibitem[{{Mart{\'\i}nez-Sykora} {et~al.}(2016){Mart{\'\i}nez-Sykora}, {De
  Pontieu}, {Hansteen}, \& {Gudiksen}}]{2016ApJ...817...46M}
{Mart{\'\i}nez-Sykora}, J., {De Pontieu}, B., {Hansteen}, V.~H., \& {Gudiksen},
  B. 2016, \apj, 817, 46

\bibitem[{{Mart{\'\i}nez-Sykora}
  {et~al.}(2017{\natexlab{b}}){Mart{\'\i}nez-Sykora}, {De Pontieu}, {Hansteen},
  {Rouppe van der Voort}, {Carlsson}, \& {Pereira}}]{Juan_2017_Science}
{Mart{\'\i}nez-Sykora}, J., {De Pontieu}, B., {Hansteen}, V.~H., {et~al.}
  2017{\natexlab{b}}, Science, 356, 1269

\bibitem[{{Mart{\'\i}nez-Sykora} {et~al.}(2011){Mart{\'\i}nez-Sykora}, {De
  Pontieu}, {Testa}, \& {Hansteen}}]{2011ApJ...743...23M}
{Mart{\'\i}nez-Sykora}, J., {De Pontieu}, B., {Testa}, P., \& {Hansteen}, V.
  2011, \apj, 743, 23

\bibitem[{{Mart{\'\i}nez-Sykora} {et~al.}(2020){Mart{\'\i}nez-Sykora},
  {Leenaarts}, {De Pontieu}, {N{\'o}brega-Siverio}, {Hansteen}, {Carlsson}, \&
  {Szydlarski}}]{2020ApJ...889...95M}
{Mart{\'\i}nez-Sykora}, J., {Leenaarts}, J., {De Pontieu}, B., {et~al.} 2020,
  \apj, 889, 95

\bibitem[{{McIntosh} {et~al.}(2007){McIntosh}, {Davey}, {Hassler}, {Armstrong},
  {Curdt}, {Wilhelm}, \& {Lin}}]{2007ApJ...654..650M}
{McIntosh}, S.~W., {Davey}, A.~R., {Hassler}, D.~M., {et~al.} 2007, \apj, 654,
  650

\bibitem[{{McIntosh} \& {De Pontieu}(2009)}]{2009ApJ...707..524M}
{McIntosh}, S.~W. \& {De Pontieu}, B. 2009, \apj, 707, 524

\bibitem[{{McIntosh} {et~al.}(2012){McIntosh}, {Tian}, {Sechler}, \& {De
  Pontieu}}]{2012ApJ...749...60M}
{McIntosh}, S.~W., {Tian}, H., {Sechler}, M., \& {De Pontieu}, B. 2012, \apj,
  749, 60

\bibitem[{{N{\'o}brega-Siverio} {et~al.}(2020){N{\'o}brega-Siverio},
  {Mart{\'\i}nez-Sykora}, {Moreno-Insertis}, \&
  {Carlsson}}]{2020A&A...638A..79N}
{N{\'o}brega-Siverio}, D., {Mart{\'\i}nez-Sykora}, J., {Moreno-Insertis}, F.,
  \& {Carlsson}, M. 2020, \aap, 638, A79

\bibitem[{{Pagano} {et~al.}(2004){Pagano}, {Linsky}, {Valenti}, \&
  {Duncan}}]{2004A&A...415..331P}
{Pagano}, I., {Linsky}, J.~L., {Valenti}, J., \& {Duncan}, D.~K. 2004, \aap,
  415, 331

\bibitem[{{Pereira} {et~al.}(2014){Pereira}, {De Pontieu}, {Carlsson},
  {Hansteen}, {Tarbell}, {Lemen}, {Title}, {Boerner}, {Hurlburt}, \&
  {W{\"u}lser}}]{Tiago_2014_heat}
{Pereira}, T.~M.~D., {De Pontieu}, B., {Carlsson}, M., {et~al.} 2014, \apj,
  792, L15

\bibitem[{{Pereira} {et~al.}(2016){Pereira}, {Rouppe van der Voort}, \&
  {Carlsson}}]{2016ApJ...824...65P}
{Pereira}, T.~M.~D., {Rouppe van der Voort}, L., \& {Carlsson}, M. 2016, \apj,
  824, 65

\bibitem[{{Pereira} {et~al.}(2018){Pereira}, {Rouppe van der Voort},
  {Hansteen}, \& {De Pontieu}}]{2018A&A...611L...6P}
{Pereira}, T. M.~D., {Rouppe van der Voort}, L., {Hansteen}, V.~H., \& {De
  Pontieu}, B. 2018, \aap, 611, L6

\bibitem[{{Pereira} \& {Uitenbroek}(2015)}]{Tiago_RH_2015}
{Pereira}, T.~M.~D. \& {Uitenbroek}, H. 2015, \aap, 574, A3

\bibitem[{{Pesnell} {et~al.}(2012){Pesnell}, {Thompson}, \&
  {Chamberlin}}]{2012SoPh..275....3P}
{Pesnell}, W.~D., {Thompson}, B.~J., \& {Chamberlin}, P.~C. 2012, \solphys,
  275, 3

\bibitem[{{Peter}(1999)}]{1999ApJ...516..490P}
{Peter}, H. 1999, \apj, 516, 490

\bibitem[{{Peter} \& {Judge}(1999)}]{1999ApJ...522.1148P}
{Peter}, H. \& {Judge}, P.~G. 1999, \apj, 522, 1148

\bibitem[{{Pneuman} \& {Kopp}(1977)}]{1977A&A....55..305P}
{Pneuman}, G.~W. \& {Kopp}, R.~A. 1977, \aap, 55, 305

\bibitem[{{Rouppe van der Voort} {et~al.}(2015){Rouppe van der Voort}, {De
  Pontieu}, {Pereira}, {Carlsson}, \& {Hansteen}}]{Luc_2015}
{Rouppe van der Voort}, L., {De Pontieu}, B., {Pereira}, T.~M.~D., {Carlsson},
  M., \& {Hansteen}, V. 2015, \apj, 799, L3

\bibitem[{{Rouppe van der Voort} {et~al.}(2009){Rouppe van der Voort},
  {Leenaarts}, {De Pontieu}, {Carlsson}, \& {Vissers}}]{Luc_2009}
{Rouppe van der Voort}, L., {Leenaarts}, J., {De Pontieu}, B., {Carlsson}, M.,
  \& {Vissers}, G. 2009, \apj, 705, 272

\bibitem[{{Rouppe van der Voort} {et~al.}(2020){Rouppe van der Voort}, {De
  Pontieu}, {Carlsson}, {de la Cruz Rodr{\'\i}guez}, {Bose}, {Chintzoglou},
  {Drews}, {Froment}, {Go{\v{s}}i{\'c}}, {Graham}, {Hansteen}, {Henriques},
  {Jafarzadeh}, {Joshi}, {Kleint}, {Kohutova}, {Leifsen},
  {Mart{\'\i}nez-Sykora}, {N{\'o}brega-Siverio}, {Ortiz}, {Pereira}, {Popovas},
  {Quintero Noda}, {Sainz Dalda}, {Scharmer}, {Schmit}, {Scullion}, {Skogsrud},
  {Szydlarski}, {Timmons}, {Vissers}, {Woods}, \&
  {Zacharias}}]{2020A&A...641A.146R}
{Rouppe van der Voort}, L.~H.~M., {De Pontieu}, B., {Carlsson}, M., {et~al.}
  2020, \aap, 641, A146

\bibitem[{{Rutten}(2020)}]{2020arXiv200900376R}
{Rutten}, R.~J. 2020, arXiv e-prints, arXiv:2009.00376

\bibitem[{{Rutten} {et~al.}(2019){Rutten}, {Rouppe van der Voort}, \& {De
  Pontieu}}]{2019A&A...632A..96R}
{Rutten}, R.~J., {Rouppe van der Voort}, L. H.~M., \& {De Pontieu}, B. 2019,
  \aap, 632, A96

\bibitem[{{Samanta} {et~al.}(2019){Samanta}, {Tian}, {Yurchyshyn}, {Peter},
  {Cao}, {Sterling}, {Erd{\'e}lyi}, {Ahn}, {Feng}, {Utz}, {Banerjee}, \&
  {Chen}}]{2019Sci...366..890S}
{Samanta}, T., {Tian}, H., {Yurchyshyn}, V., {et~al.} 2019, Science, 366, 890

\bibitem[{{Scharmer} {et~al.}(2003){Scharmer}, {Bjelksj{\"o}}, {Korhonen},
  {Lindberg}, \& {Petterson}}]{2003SPIE.4853..341S}
{Scharmer}, G.~B., {Bjelksj{\"o}}, K., {Korhonen}, T.~K., {Lindberg}, B., \&
  {Petterson}, B. 2003, in Society of Photo-Optical Instrumentation Engineers
  (SPIE) Conference Series, Vol. 4853, Innovative Telescopes and
  Instrumentation for Solar Astrophysics, ed. S.~L. {Keil} \& S.~V. {Avakyan},
  341--350

\bibitem[{{Scharmer} {et~al.}(2019){Scharmer}, {L{\"o}fdahl}, {Sliepen}, \& {de
  la Cruz Rodr{\'\i}guez}}]{2019A&A...626A..55S}
{Scharmer}, G.~B., {L{\"o}fdahl}, M.~G., {Sliepen}, G., \& {de la Cruz
  Rodr{\'\i}guez}, J. 2019, \aap, 626, A55

\bibitem[{{Scharmer} {et~al.}(2008){Scharmer}, {Narayan}, {Hillberg}, {de la
  Cruz Rodriguez}, {L{\"o}fdahl}, {Kiselman}, {S{\"u}tterlin}, {van Noort}, \&
  {Lagg}}]{Crisp_2008}
{Scharmer}, G.~B., {Narayan}, G., {Hillberg}, T., {et~al.} 2008, \apj, 689, L69

\bibitem[{{Sekse} {et~al.}(2012){Sekse}, {Rouppe van der Voort}, \& {De
  Pontieu}}]{Sekse_2012}
{Sekse}, D.~H., {Rouppe van der Voort}, L., \& {De Pontieu}, B. 2012, \apj,
  752, 108

\bibitem[{{Sekse} {et~al.}(2013){Sekse}, {Rouppe van der Voort}, {De Pontieu},
  \& {Scullion}}]{2013ApJ...769...44S}
{Sekse}, D.~H., {Rouppe van der Voort}, L., {De Pontieu}, B., \& {Scullion}, E.
  2013, \apj, 769, 44

\bibitem[{{Skogsrud} {et~al.}(2014){Skogsrud}, {Rouppe van der Voort}, \& {De
  Pontieu}}]{2014ApJ...795L..23S}
{Skogsrud}, H., {Rouppe van der Voort}, L., \& {De Pontieu}, B. 2014, \apjl,
  795, L23

\bibitem[{{Skogsrud} {et~al.}(2016){Skogsrud}, {Rouppe van der Voort}, \& {De
  Pontieu}}]{2016ApJ...817..124S}
{Skogsrud}, H., {Rouppe van der Voort}, L., \& {De Pontieu}, B. 2016, \apj,
  817, 124

\bibitem[{{Sukhorukov} \& {Leenaarts}(2017)}]{2017A&A...597A..46S}
{Sukhorukov}, A.~V. \& {Leenaarts}, J. 2017, \aap, 597, A46

\bibitem[{{Tian} {et~al.}(2014){Tian}, {DeLuca}, {Cranmer}, {De Pontieu},
  {Peter}, {Mart{\'\i}nez-Sykora}, {Golub}, {McKillop}, {Reeves}, {Miralles},
  {McCauley}, {Saar}, {Testa}, {Weber}, {Murphy}, {Lemen}, {Title}, {Boerner},
  {Hurlburt}, {Tarbell}, {Wuelser}, {Kleint}, {Kankelborg}, {Jaeggli},
  {Carlsson}, {Hansteen}, \& {McIntosh}}]{2014Sci...346A.315T}
{Tian}, H., {DeLuca}, E.~E., {Cranmer}, S.~R., {et~al.} 2014, Science, 346,
  1255711

\bibitem[{{Uitenbroek}(2001)}]{Uitenbroek_2001}
{Uitenbroek}, H. 2001, \apj, 557, 389

\bibitem[{{van der Velden}(2020)}]{2020JOSS....5.2004V}
{van der Velden}, E. 2020, The Journal of Open Source Software, 5, 2004

\bibitem[{{van Noort} {et~al.}(2005){van Noort}, {Rouppe van der Voort}, \&
  {L{\"o}fdahl}}]{vannoort2005MOMFBD}
{van Noort}, M., {Rouppe van der Voort}, L., \& {L{\"o}fdahl}, M.~G. 2005,
  \solphys, 228, 191

\bibitem[{{Vissers} \& {Rouppe van der Voort}(2012)}]{2012ApJ...750...22V}
{Vissers}, G. \& {Rouppe van der Voort}, L. 2012, \apj, 750, 22

\bibitem[{{Wood} {et~al.}(1996){Wood}, {Harper}, {Linsky}, \&
  {Dempsey}}]{1996ApJ...458..761W}
{Wood}, B.~E., {Harper}, G.~M., {Linsky}, J.~L., \& {Dempsey}, R.~C. 1996,
  \apj, 458, 761

\bibitem[{{Zacharias} {et~al.}(2018){Zacharias}, {Hansteen}, {Leenaarts},
  {Carlsson}, \& {Gudiksen}}]{2018A&A...614A.110Z}
{Zacharias}, P., {Hansteen}, V.~H., {Leenaarts}, J., {Carlsson}, M., \&
  {Gudiksen}, B.~V. 2018, \aap, 614, A110

\end{thebibliography}

\end{document}